\documentclass[twocolumn,prd,noshowpacs,nofootinbib,amsmath,amssymb]{revtex4}
\usepackage{graphicx,epsfig,psfrag,bm,amssymb}
\usepackage{dcolumn,float}
\usepackage{bm}
\usepackage{color}
\usepackage{mathrsfs,amsfonts,hepunits, color}
\usepackage{mciteplus}
\usepackage{tikz}
\RequirePackage{xspace}
\usepackage{relsize}
\usepackage{slashed}

\newcommand{\be}{\begin{eqnarray}}
\newcommand{\ee}{\end{eqnarray}}
\def\lsim{\mathrel{\rlap{\lower4pt\hbox{\hskip 0.5 pt$\sim$}}
\raise1pt\hbox{$<$}}}                % less than or approx. symbol
\def\gsim{\mathrel{\rlap{\lower4pt\hbox{\hskip1pt$\sim$}}
\raise1pt\hbox{$>$}}} 
    
\newcommand{\apr}{{A^\prime}}

\begin{document}

\title{Physics Motivation for a Pilot Dark Matter Search at Jefferson Laboratory 
}  
\author{Eder Izaguirre, Gordan Krnjaic, Philip Schuster, and Natalia Toro}

\vspace{2cm}
\affiliation{ \\ Perimeter Institute for Theoretical Physics, Waterloo, Ontario, Canada   N2L 2Y5 }
\vspace{5cm}
\date{\today}

\begin{abstract}
It has recently been demonstrated that a program of parasitic
electron-beam fixed-target experiments would have powerful discovery
potential for dark matter and other new weakly-coupled particles in
the MeV--GeV mass range. The first stage of this program can be
realized at Jefferson Laboratory using an existing
plastic-scintillator detector downstream of the Hall D electron beam
dump.  This paper studies the physics potential of such an experiment
and highlights its unique sensitivity to inelastic ``exciting'' dark matter and leptophilic dark matter scenarios. The first of these is kinematically
inaccessible at traditional direct detection experiments and features
potential ``smoking gun'' low-background signatures. 
\end{abstract}

\maketitle

%%%%%%%%%%%%%%%%%%%%%%%%%%%%%%%%%%%%%%%%%%%%%%%%%%%%
%%%%%%%%%%%%%%%%%%%%%%%%%%%%%%%%%%%%%%%%%%%%%%%%%%%%
%
%								SEC: Introduction
%
%%%%%%%%%%%%%%%%%%%%%%%%%%%%%%%%%%%%%%%%%%%%%%%%%%%%
%%%%%%%%%%%%%%%%%%%%%%%%%%%%%%%%%%%%%%%%%%%%%%%%%%%%

\section{Introduction}

Although overwhelming astrophysical and cosmological evidence supports
the existence of dark matter (DM) \cite{Bergstrom:2012fi}, its
identity, interactions, and origin remain elusive.  There is currently
an active program to probe particle DM scattering with direct detection
experiments, annihilation with indirect detection telescopes, and
production with particle accelerators \cite{Beringer:1900zz}. 
However, most of these efforts are designed to find
heavy (10$-$1000 GeV) DM candidates and sharply lose sensitivity to
lighter (sub-GeV) states whose signals are either too feeble or lie in
high-background regions. Even direct-detection experiments
\cite{Agnese:2013jaa, Barreto:2011zu, Essig:2012yx} and proposals \cite{Essig:2011nj, Graham:2012su, Gerbier:2014jwa} that are expanding
sensitivity to GeV-scale DM rely on an elastic scattering channel that
is absent or highly suppressed in many DM scenarios  \cite{Finkbeiner:2014sja,TuckerSmith:2001hy,Finkbeiner:2007kk,Chang:2008gd,Cui:2009xq,Finkbeiner:2009mi,Chang:2010en,Pospelov:2013nea}.

%%%%%%%%%%%%%%%%%%%%%%%%%%%%%%%%%%%%%%%%%%%%%%%%
% 					               Scattering Figure 
%%%%%%%%%%%%%%%%%%%%%%%%%%%%%%%%%%%%%%%%%%%%%%%%

\begin{figure}[t!]
\includegraphics[width=8.6cm]{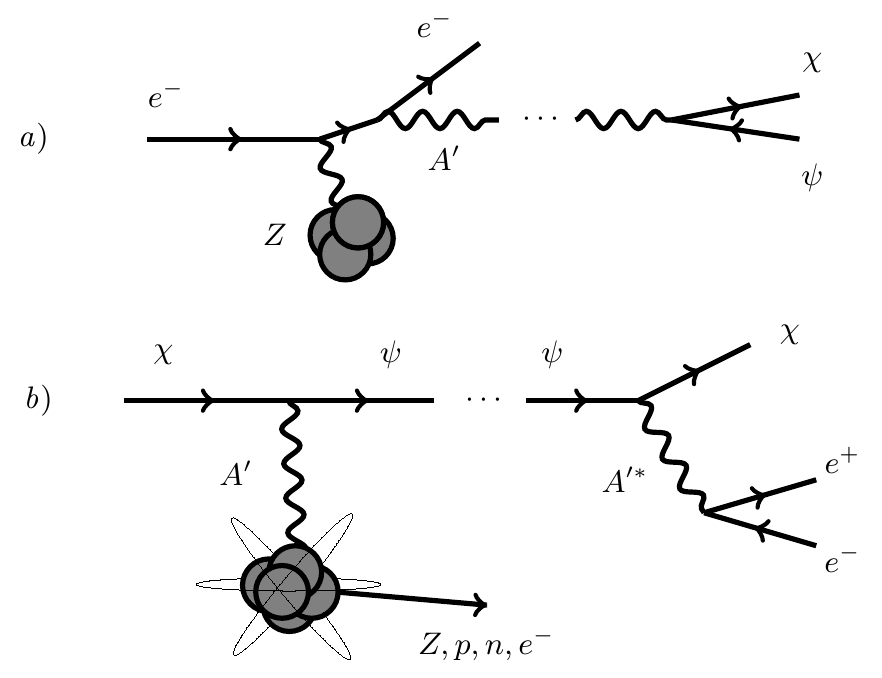}
\caption{  a) Fermionic DM pair production from $\apr$-sstrahluung in electron-nucleus collisions.
In the generic scenario with Dirac and Majorana masses for dark sector fermions,
the $\apr$ mediator couples off diagonally to the mass eigenstates $\chi$ and $\psi$ (see Sec. \ref{sec:fermions}). 
 b) Detector scattering via $A^\prime$ exchange inside the detector. If the mass splitting between
  dark sector states is negligible, both the incoming and outgoing DM states in the 
  scattering process are invisible and can be 
  treated as the same particle.  For order one (or larger) mass splittings, $\chi$ can upscatter into 
  the excited state $\psi$, which promptly decays inside the detector via
 $\psi\to\chi\, e^+e^-$. This  process yields a target (nucleus, nucleon, or electron) recoil $E_R$ 
 and two charged tracks, which is a distinctive, low background signature, so nuclear recoil cuts need
 not be limiting. Processes analogous to both a) and b) can also exist if DM is a scalar -- see Sec. \ref{sec:scalars}
 }
\label{fig:scatfig}  
\end{figure}

Recently it was shown that electron-beam fixed target experiments offer powerful sensitivity to a broad 
class of dark sector scenarios that feature particles in the elusive MeV-GeV mass range \cite{Izaguirre:2013uxa, Diamond:2013oda}. 
If DM couples to leptonic currents via mediators of comparable mass, it can be produced copiously 
 in relativistic electron-nucleus collisions and scatter in a downstream detector (see Fig. \ref{fig:scatfig}). 
Electron beam-dump experiments are complementary to dedicated efforts at proton beam facilities 
 \cite{Batell:2009di,deNiverville:2011it, deNiverville:2012ij,Dharmapalan:2012xp,Essig:2013lka}, and have comparable DM scattering yield. 
Electron-beam experiments can run
parasitically on a smaller scale and benefit from negligible beam-related backgrounds.

Jefferson Laboratory (JLab) is currently  upgrading 
its 6 GeV electron beam to operate at 12 GeV energies. 
The new CEBAF (continuous electron beam accelerator facility) is scheduled to begin delivering $\sim 100 \mu$A 
currents in mid-2014 and presents new opportunities to search
for new light weakly coupled particles.  
A possible first step would be a parasitic pilot experiment using an existing plastic-scintillator detector behind the Hall D electron beam dump, which will receive a $\sim 200$ nA current \cite{testProposal}.  Such an experiment could pave the way for a larger-scale experiment behind a higher-current beam dump \cite{Izaguirre:2013uxa}.  Remarkably, even a small-scale pilot experiment has potential discovery sensitivity to several DM scenarios, which we explore in this paper.  A particularly dramatic signal could be seen if DM states are split by $\gtrsim$ MeV, so that DM scattering produces energetic $e^+e^-$ pairs (considered in other contexts in \cite{Finkbeiner:2014sja,Finkbeiner:2007kk,Pospelov:2013nea,Morris:2011dj,Vincent:2012an,Finkbeiner:2009mi,Pospelov:2007xh,Cholis:2008vb,Batell:2009vb}).

The basic production and detection processes we consider here parallel those discussed in \cite{Batell:2009di,deNiverville:2011it,Izaguirre:2013uxa}. Electrons impinging on atomic
nuclei in a beam dump can emit light mediator particles that promptly
decay to pairs of DM particles or the DM can be radiated via off shell
mediator exchange (Figure \ref{fig:scatfig}(a)). 
%For a thorough discussion of production rates, see \cite{Izaguirre:2013uxa}. 
The pair of DM particles emerge from the beam dump in a highly collimated beam and pass through
the shielding and dirt because their interactions are weak.  A fraction of the DM particles scatter off electrons, nucleons, or nuclei via mediator exchange  in a downstream detector (Figure \ref{fig:scatfig}(b), left).  Because the DM particles are relativistic, their scattering can induce multi-MeV recoils of the target which in turn produce scintillation or \v Cerenkov light.

Our treatment generalizes \cite{Izaguirre:2013uxa} in three important ways.  First, we consider the possibility that the mediator coupling DM to SM matter couples \emph{only} to leptons, not to nucleons --- for example, a vector can couple to the conserved $U(1)_{e-\mu}$ current.  This scenario produces only electron-scattering, but no associated nucleon/nucleus scattering signal.  In this case, DM would not be produced in proton-beam experiments,  but neutrino physics does constrain the $U(1)_{e-\mu}$ coupling.  Second, whereas \cite{Izaguirre:2013uxa}  focused primarily on quasi-elastic scattering off nucleons, we consider DM-electron, DM-nucleon, and DM-nucleus scattering here.  The latter is most significant at low momentum transfers (where it is $Z^2$-enhanced), but is suppressed by form factors at higher momentum transfers.   DM-electron scattering can easily yield multi-GeV electron recoils for the mediator masses of interest, and are therefore particularly visible.  Third, and most significantly, we consider a new signal that arises when the DM states have $O(1)$ mass splittings and have appreciable inelastic interactions, as in  \cite{Finkbeiner:2014sja,TuckerSmith:2001hy,Finkbeiner:2007kk,Chang:2008gd,Cui:2009xq,Finkbeiner:2009mi,Chang:2010en,Pospelov:2013nea}. In this case, the up-scattering of a dark matter state $\chi$ into an excited state $\psi$ is followed by a prompt decay $\psi \rightarrow \chi e^+e^-$ (Figure \ref{fig:scatfig}(b), right).  Thus, the target recoil is accompanied by the GeV-scale energy deposition of the $e^+e^-$ pair, which can carry a significant fraction of the incident beam energy. Because beam-related backgrounds are small and cosmic backgrounds are dominated at much lower energies, the energy deposition from such decays could be a ``smoking gun'' signal for a light DM candidate even in a small above-ground detector.

The outline of this paper is as follows. Section \ref{sec:models} presents the simplified models we consider in our analyses and discusses the model dependence of existing constraints. Section \ref{sec:testrun} describes the setup of a test run at JLab Hall D (inspired by \cite{MarcoSlides}) and 
presents our yield projections for different scattering channels. Finally, section \ref{sec:conclusion} 
offers some concluding remarks.
 
%%%%%%%%%%%%%%%%%%%%%%%%%%%%%%%%%%%%%%%%%%%%%%%%
%%%%%%%%%%%%%%%%%%%%%%%%%%%%%%%%%%%%%%%%%%%%%%%%
%
% 					        SEC: Review DM production & detection
%
%%%%%%%%%%%%%%%%%%%%%%%%%%%%%%%%%%%%%%%%%%%%%%%%
%%%%%%%%%%%%%%%%%%%%%%%%%%%%%%%%%%%%%%%%%%%%%%%%

%%%%%%%%%%%%%%%%%%%%%%%%%%%%%%%%%%%%%%%%%%%%%%%%
%%%%%%%%%%%%%%%%%%%%%%%%%%%%%%%%%%%%%%%%%%%%%%%%
%
% 					              SEC:  Benchmark Scenarios
%
%%%%%%%%%%%%%%%%%%%%%%%%%%%%%%%%%%%%%%%%%%%%%%%%
%%%%%%%%%%%%%%%%%%%%%%%%%%%%%%%%%%%%%%%%%%%%%%%%

\section{Benchmark Scenarios} 
\label{sec:models}

\begin{figure}[h!]
 \vspace{0.1cm}
\includegraphics[width=8.3cm]{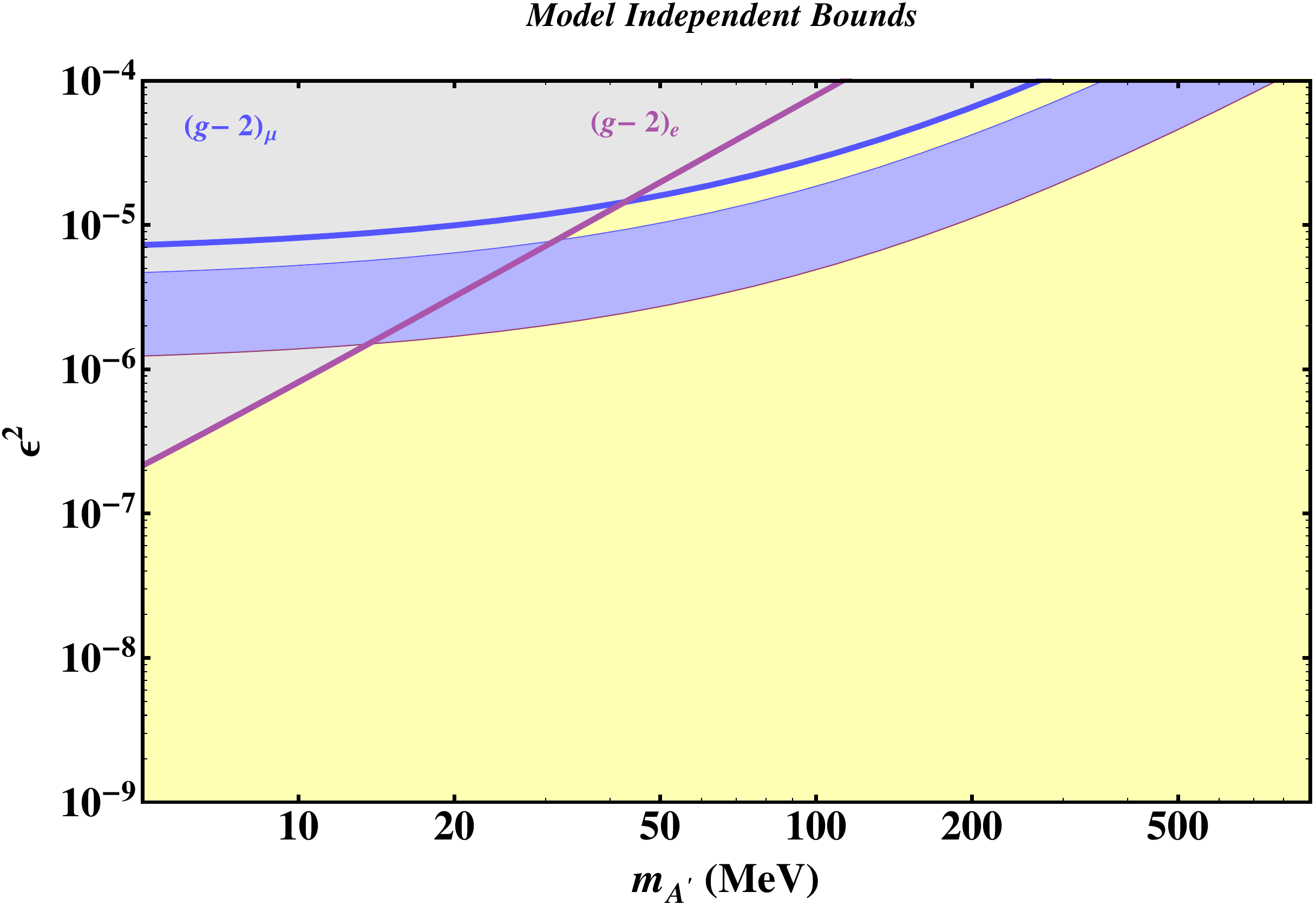}
\includegraphics[width=8.3cm]{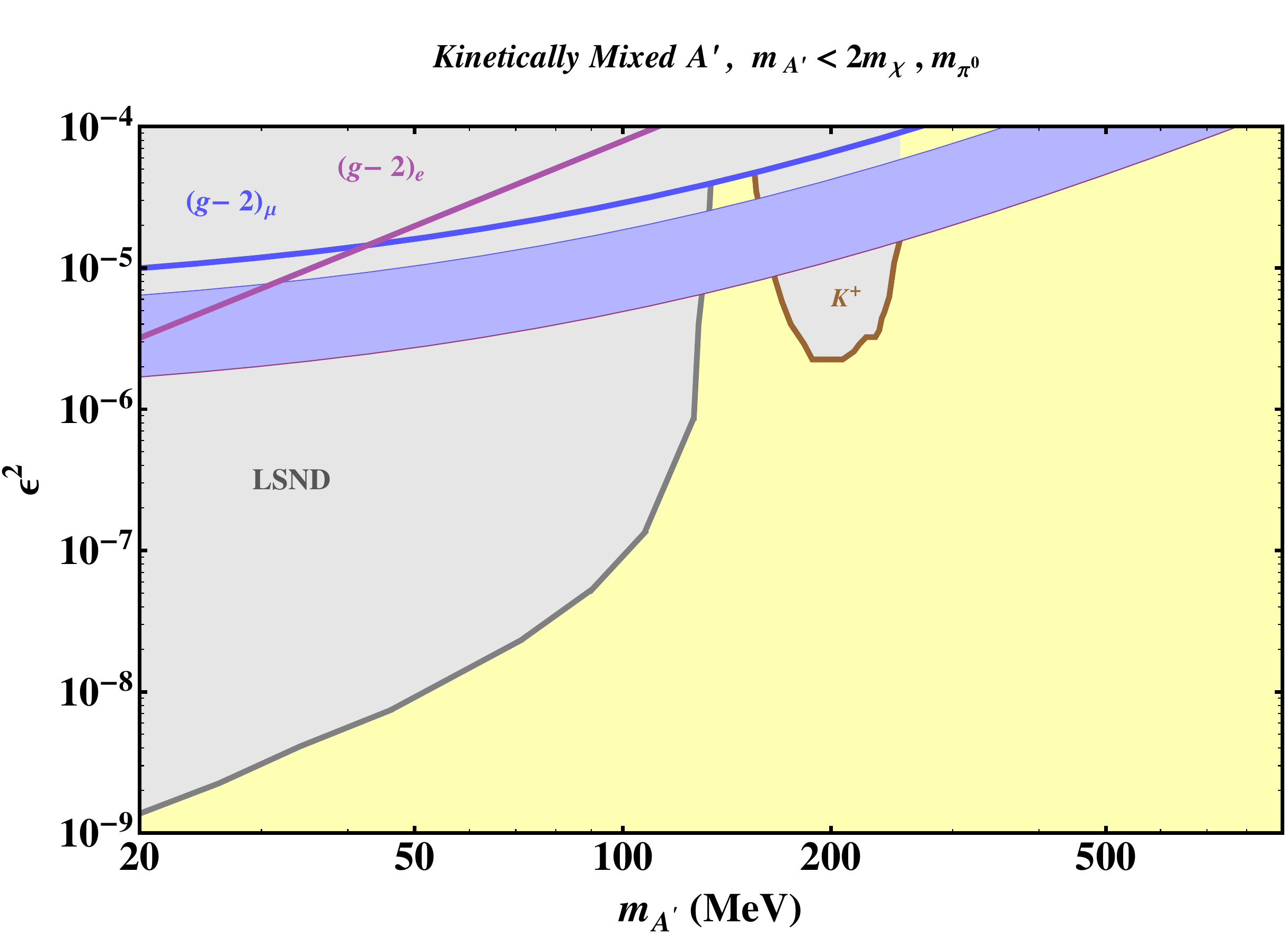}
\includegraphics[width=8.3cm]{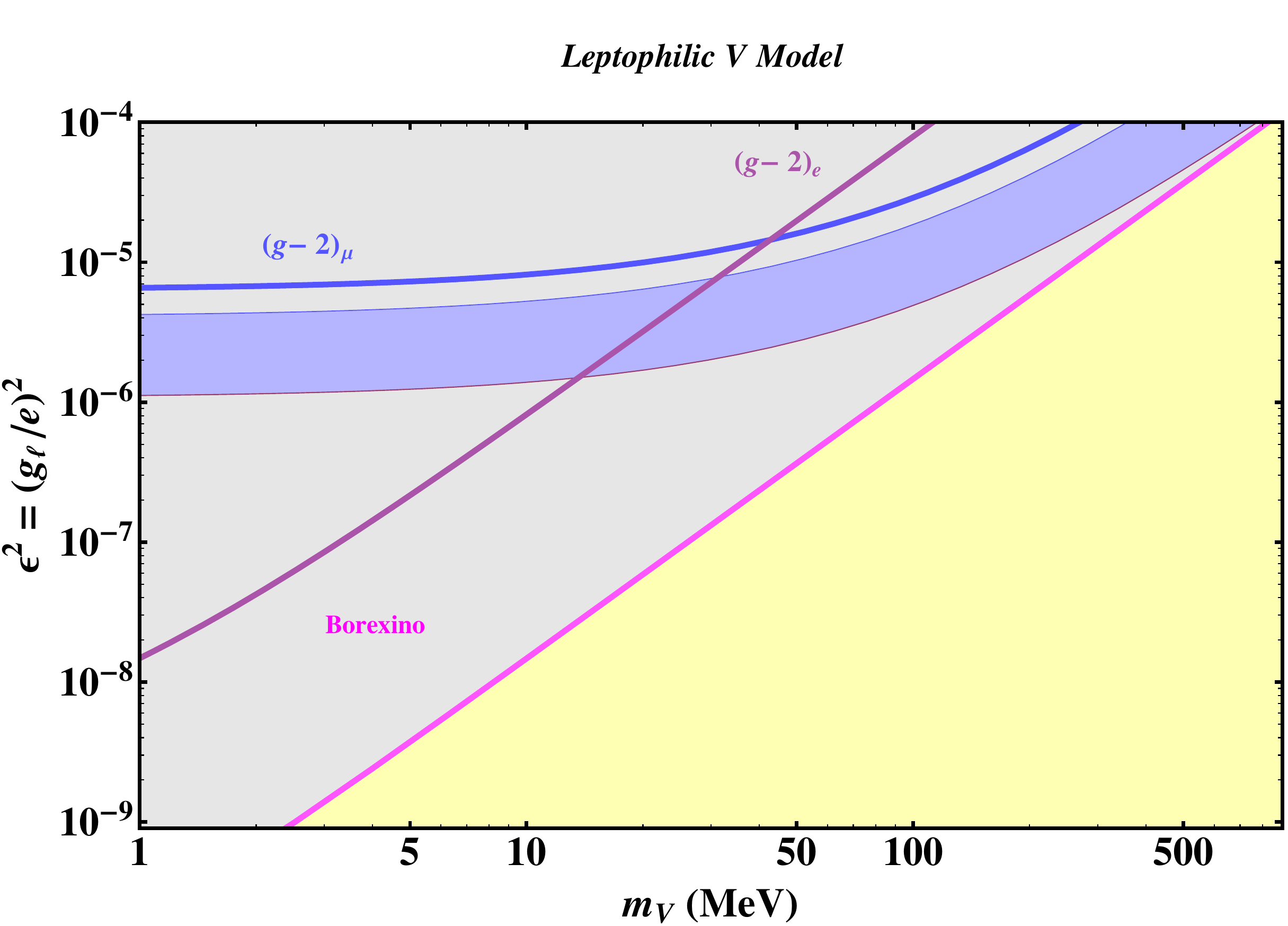}
  \caption{
Top: model independent bounds on an MeV-GeV scae gauge boson that couples only to charged leptons.  
The purple band is the region that resolves $(g-2)_\mu$ anomaly to within $2\sigma$ \cite{Pospelov:2008zw}.  
Middle: model {\it dependent} bounds on a kinetically 
mixed $\apr$ from Sec. \ref{sec:aprime-mediator} assuming it 
decays predominantly to two stable, invisible fermions in the dark sector $\apr \to \chi\bar\chi$ and $m_\apr  \gg 2 m_\chi$ 
as outlined in Sec. \ref{sec:fermions}. Additional constraints are from LSND \cite{deNiverville:2011it}, a mono-$\gamma$ search
 at BaBar \cite{Aubert:2008as}, and rare Kaon decays \cite{Artamonov:2008qb}. Bottom: bounds on the leptophilic model from
  Sec. \ref{sec:leptophilic-mediator}. This model requires $V$ couplings to neutrinos which are constrained by
 Borexino \cite{Bellini:2011rx}.   
}
\label{fig:depindep}
\end{figure}

Viable MeV--GeV-mass dark matter candidates that thermalize in the early Universe require a light mediator through which the DM can annihilate.  
Surprisingly,  there are few {\it model independent} constraints on
light, leptonically coupled mediators in the MeV--GeV mass range (see \cite{Essig:2013lka}  for a 
review). Fig.~\ref{fig:depindep} (top) shows the 
bounds on a vector mediator from precision QED measurements without assuming 
anything about its decay or scattering  signatures. We require  only that the mediator 
 couples to leptons with strength $\epsilon e$; insertions of virtual $\apr$ into diagrams
 generically correct the lepton-photon vertex and contribute to $(g-2)_{e,\mu}$.  
A growing program of direct searches for light mediators rely either on visible decays of these mediators \cite{Bjorken:2009mm,Bjorken:1988as,Riordan:1987aw,Bross:1989mp,Essig:2009nc,Essig:2010xa,Batell:2009yf,Fayet:2007ua,Freytsis:2009bh,Essig:2010gu,Reece:2009un,Wojtsekhowski:2009vz,AmelinoCamelia:2010me,Baumgart:2009tn,Merkel:2011ze,Beranek:2013yqa,Abrahamyan:2011gv,Aubert:2009cp,Echenard:2012hq,Babusci:2012cr,Adlarson:2013eza,Davier:1986qq,Hook:2010tw,Morrissey:2014yma} or their hadronic couplings \cite{Batell:2009di,deNiverville:2011it, deNiverville:2012ij,Dharmapalan:2012xp,Essig:2013lka}.
In this section we consider the constraints on a variety of new mediators
 and establish benchmark DM scenarios for which
electron beam dump experiments are particularly sensitive. 

%%%%%%%%%%%%%%%%%%%%%%
%		  FIG  Generic Stuff
%%%%%%%%%%%%%%%%%%%%%%

%%%%%%%%%%%%%%%%%%%%
%			 Mediators to the SM 
%%%%%%%%%%%%%%%%%%%%

\subsection{Mediators to the Standard Model}
\label{sec:mediators}
We consider two representative possibilities for the mediator interactions with the Standard Model: kinetic mixing with the photon (which couples to all charged leptons and hadrons) and interactions with leptons alone through a $U(1)_{e-\mu}$ coupling.  Up to numerical factors, results for the kinetic mixing model can be taken as a proxy for any mediator whose interactions with electrons and light nucleons have comparable strengths.  The $U(1)_{e-\mu}$ model is likewise representative of the more general possibility that the mediator interacts with electrons but \emph{not} nucleons.  

% Kinetically Mixed U(1)_D

\subsubsection{Kinetically Mixed $U(1)_D$}
\label{sec:aprime-mediator}
For simplicity, 
it is convenient to frame our discussion in terms a simplified model with a massive, 
{\it invisibly decaying} vector boson $A^\prime$ from a broken $U(1)_D$ 
gauge group. The most general lagrangian  for $\apr$ contains
\be \label{eq:aprime-lagrangian}
 {\cal L}  \supset   
  - \frac{1}{4}{F^\prime}^{\mu\nu} F^\prime_{\mu\nu}
   +\frac{\epsilon}{2}F^{\mu\nu} F^\prime_{\mu\nu} + \frac{m^2_{V}}{2} A^{\prime}_\mu A^{\prime \mu} ~~,
\ee
where $F^\prime_{\mu\nu} \equiv \partial_{\,[\mu, } \apr_{\nu]}$  is the field strength,
$m_A^\prime \sim$ MeV -- GeV is the mediator mass, and 
$\epsilon$ is the kinetic mixing, which is naturally in the $10^{-5} - 10^{-2}$ range if generated by loops
of heavy particles charged under both dark and visible gauge groups. 
Diagonalizing the gauge kinetic terms induces an effective coupling 
to SM currents 
\be\label{eq:aprime-currents}
{\cal L} \supset \epsilon  \apr_\mu \sum_i q_i \overline f_i  \gamma^\mu f_i~~,
\ee
where $f_i$ is any SM quark or lepton and $q_i$ is its charge. Assuming the
$\apr$ decays invisibly\footnote{Constraints on a visibly decaying mediator 
 $\apr \to e^+ e^-$ are also given in \cite{Essig:2010ye}, but we do not consider this scenario.}
  to pairs of dark-sector states with masses below $\sim$ 68 MeV, this extension faces constraints from BaBar \cite{Izaguirre:2013uxa, Essig:2013vha},
   LSND \cite{deNiverville:2011it},  $(g-2)_{e,\mu}$ \cite{Giudice:2012ms, Pospelov:2008zw},
   and  rare Kaon decays \cite{Artamonov:2008qb}, shown in the bottom panel of Fig.~\ref{fig:depindep}.

The interactions in Eq.~(\ref{eq:aprime-currents}) mediate DM scattering off electrons, coherent scattering off nuclei, quasielastic scattering off nucleons, and inelastic scattering off nuclei. The last process requires substantial momentum transfer and is not included in our simulations. For a detailed discussion of the signals we simulate in our numerical studies, see Sec.~\ref{sec:simulation} and Appendix A.

%Leptophilic Mediator 
\subsubsection{Leptophilic $U(1)_{e-\mu}$   }
\label{sec:leptophilic-mediator}
The simplest leptonically coupled mediator arises from a $U(1)_{\ell_i - \ell_j}$ gauge 
extension to the SM \cite{Fox:2008kb,Cirelli:2008pk,Chen:2008dh,Baltz:2002we,Agrawal:2014ufa,Chang:2014tea}, where $\ell_{i,j} = e, \mu,$ or $ \tau$ and $i\ne j$ are SM
leptons. For concreteness, we consider only $U(1)_{e-\mu}$ as the simplest model that allows mediator couplings to 
electrons.  The lagrangian for this mediator contains 
\be \label{eq:leptophilic-lagrangian}
 {\cal L}  \supset   
  - \frac{1}{4}{\cal F}^{\mu\nu} {\cal F}_{\mu\nu} +
    \frac{m^2_{V}}{2}  { V}_\mu { V}^{\mu} +  { V}_\mu         \sum_i g_{\ell_i} \overline \ell_i \gamma^\mu \ell_i ~~,
\ee
where ${\cal F_{\mu \nu}} =  \partial_{[\mu,} V_{\nu]}$ is the field strength
 and $g_{\ell_i}$ is the $U(1)_{e-\mu}$ charge for lepton $\ell_i$. 

In this mass range leptophilic invisibly decaying mediators are constrained only by precision QED measurements of 
 $(g-2)_{e,\mu}$ \cite{Giudice:2012ms, Pospelov:2008zw} and by neutrino-scattering observations with Borexino \cite{Bellini:2011rx}. For comparison with 
 the conventional bounds on kinetically mixed gauge bosons, we will 
 present the parameter space in terms of the parameter $\epsilon \equiv g_\ell / e$. 
 
%%%%%%%%%%%%%%%%%%%%%%%%%%%%%%%
%			 Dark Sector Spectra
%%%%%%%%%%%%%%%%%%%%%%%%%%%%%%%

\subsection{Dark Species and Spectra}
\label{sec:darkside}
We now consider simplified models of the dark sector that feature either a  complex scalar or Dirac
fermion coupled to the SM via one of the mediators in Sec.~\ref{sec:mediators}.  For simplicity, in this subsection we 
 use the notation appropriate for the $U(1)_D$ model with an $\apr$, but the features discussed below apply equally to a $U(1)_{e-\mu}$ gauge boson or any other spin-1 mediator.  

A key feature of these models is that the same spontaneous symmetry breaking that gives the mediator a non-zero mass (for concreteness, we consider a perturbative Higgs mechanism) can also split the bosonic/fermionic matter into two real/Majorana states with different masses.  The leading mediator coupling in these cases are \emph{generically} off-diagonal. Thus the DM production mode shown in Figure \ref{fig:scatfig}(a) always produces one light and one heavy particle, and its scattering (Figure \ref{fig:scatfig}(b) left) is always inelastic.  
The subsequent phenomenology is determined by the excited-state lifetime, which scales as $m_{A'}^4/(\Delta^{5} \epsilon^2 \alpha \alpha_D)$ and so is very sensitive to the size of the splitting.  For large enough splittings, the decay occurs inside the detector and the decay products contribute significantly to (or even dominate) the energy deposition from DM scattering (see \S\ref{sec:smoking}).
These inelastic scenarios are especially important to consider because a thermal origin for dark matter in such models is entirely compatible with constraints on light dark matter derived from measurements of the CMB \cite{toappear}. 
 
%%%%%%%%%%%%%%%%%%%%%%%%%%%%%%%%%%%%%%%%%%%%%%%
%					Scalar Spectrum
%%%%%%%%%%%%%%%%%%%%%%%%%%%%%%%%%%%%%%%%%%%%%%%%
\subsubsection{Scalar Spectra}
\label{sec:scalars}
Consider a complex scalar particle $\Phi$ coupled to a $U(1)_D$ gauge boson that gets its mass from the symmetry-breaking vev of a second charged scalar, $H_D$.  If $\Phi$ and $H_D$ have equal and opposite charges, the most general Lagrangian contains 
\be\label{eq:scalar-lag}
{\cal L}\supset |(\partial_\mu - i g_D A_\mu) \Phi|^2 & -& \left( M^2 +  \eta|H_D|^2 \right) |\Phi|^2 \nonumber \\
& - & \kappa \Phi^2 H_D^2 -  \lambda | \Phi |^4 + h.c.,
\ee
where $g_D = \sqrt{4\pi \alpha_D}$ is the $U(1)_D$ coupling constant to \emph{dark sector} matter.
For $\langle H_D\rangle \ne 0$, the potential contains both diagonal and  
off-diagonal mass terms for $\Phi$, which split the mass-eigenstates. The mass eigenbasis now
features two states $\varphi$ and $\phi$ whose mass splitting $M_\phi - M_\varphi \equiv \Delta$ is generically of order the common
mass scale in the dark sector (or smaller for small $\kappa$). 

After symmetry breaking, the mass eigenstates couple off-diagonally to the mediator, 
via the derivative interaction $g_D \apr_\mu \phi i\partial^\mu \varphi + c.c.$. Thus, in the presence of mass splittings, 
every $\apr$ produced in a beam dump yields a ground state $\varphi$ and an excited state $\phi$, which can
generate distinct detector signatures. An incident excited state $\phi$ scatters by converting into the ground state
 and inelastically depositing it energy into the target particle via $\apr$ exchange. The ground state $\varphi$ can only interact with the detector by upscattering into the excited state, which for $\Delta < m_\apr$  
decays via $\phi \to  \apr^* \to \varphi e^+e^-$.  In the  $\Delta\ll  M_\phi$ limit, the width for this process is   
\be
 \label{eq:scalar-width}
\Gamma(\phi \to \varphi ~e^+e^-)  = \frac{4\epsilon^2 \alpha \alpha_D \Delta^5}{15 \pi  m_{A^\prime}^4} + {\cal O}(\Delta^6)~~,~~
\ee
(see Appendix B). 
For a boost factor of $\gamma$, the decay length is
\be
\label{eq:decay-length}
\ell_\phi  &=&  \gamma c / \Gamma(\phi \to \varphi ~e^+e^-)  \\   \nonumber \\&&  \hspace{-1cm}\simeq  0.01 \cm 
\left(    \frac{\gamma}{2} \right) 
 \left(\frac{10^{-3}}{\epsilon} \right)^2  \! 
 \left(    \frac{0.1}{\alpha_D} \right)  \!
  \left(    \frac{50 ~\MeV}{\Delta} \right)^{\!5}  \!
  \left(    \frac{m_\apr}{50 ~\MeV} \right)^{\!4} ,  \nonumber 
\ee
so for splittings of order the mediator mass, the decay is microscopic on detector length scales and gives rise to a 
distinctive signal. 

%%%%%%%%%%%%%%%%%%%%%%%%%%%%%%%%%%%%%%%%%%%%%%%
%					Fermion Spectrum
%%%%%%%%%%%%%%%%%%%%%%%%%%%%%%%%%%%%%%%%%%%%%%%%
\subsubsection{Fermionic Spectra} 
\label{sec:fermions}

If the $\apr$ interacts with a dark sector Dirac fermion $\Psi = (\lambda, \xi^\dagger)$ charged under $U(1)_D$, the Lagrangian 
in Weyl components is
\be
{\cal L} \supset  i\lambda^\dagger \bar \sigma^\mu D_\mu  \lambda + i\xi^\dagger \bar \sigma^\mu D_\mu \xi + m (\xi \lambda + \xi^\dagger \lambda^\dagger) + h.c.
\ee
Again, 
 for appropriate charge assignments, there are also Yukawa interactions
\be
{\cal L} \supset y_\lambda H_D \lambda \lambda +  y_\xi H_D^\dagger \xi \xi + h.c., 
\ee
which induce Majorana mass terms after spontaneous symmetry breaking. Diagonalizing the 
fermion masses yields states $\psi$ and $\chi$ with masses $\sim m \pm y \langle H_D \rangle$, respectively and the gauge 
mediator couples off diagonally to the mass eigenstates via  the $g_D \apr_\mu \psi^\dagger \bar \sigma^\mu \chi + h.c.$ interaction. 

Scattering through $\apr$ exchange is now necessarily inelastic and the heavier state $\psi$ can now de-excite (see Appendix B) with width 
\be
 \label{eq:fermion-width}
\Gamma(\psi \to \chi ~e^+e^-)  = \frac{8 \epsilon^2 \alpha \alpha_D \Delta^5 }{15 \pi m_\apr^4 } + {\cal O}(\Delta^6),  
\ee
which is parametrically comparable to the corresponding scalar result in Eq.~(\ref{eq:scalar-width}).

%%%%%%%%%%%%%%%%%%%%%%%%%%%%%%%%%%%%%%%%%%%%%%%%
%					Smoking Gun Signatures
%%%%%%%%%%%%%%%%%%%%%%%%%%%%%%%%%%%%%%%%%%%%%%%%
\subsection{Smoking Gun Signals}
\label{sec:smoking}

In \cite{Izaguirre:2013uxa} it was shown that electron beam dumps have sensitivity to
quasi-elastic DM-nucleon scattering,  $\chi n \to \chi n$, via $\apr$ exchange, however, 
for continuous wave (CW) beams
\footnote{For a  CW beam, the beam-on live-time coincides with the total duration of the experiment, which
is on the order of several-months, so timing is difficult and the detector encounters the maximum 
flux of environmental backgrounds over that time interval.  
In contrast, a {\it pulsed} beam delivers electrons  
in small, concentrated bunches, so the beam-on time is  typically $\ll 10^{-2}$ of the total 
experimental run, which dramatically reduces the detector's effective exposure to environmental backgrounds.},
 exploiting this process typically requires shielding or vetoing environmental backgrounds. 
However, there are two {\it smoking 
gun} signals with such high energy deposition that the backgrounds can be dramatically reduced or even eliminated (see \S\ref{sec:backgrounds}); even a $\sim 10$-event signal of these types could suffice for a convincing discovery. 

\subsubsection{High Energy Electron Recoils}
For both mediator models in Sec.~\ref{sec:mediators} and both dark sector scenarios in Sec.~\ref{sec:darkside}, a typical
DM particle produced in the beam dump can scatter off detector electrons and produce
visible recoil energies. The dominant backgrounds for this channel comes from cosmic muons which either decay in flight or are stopped in the detector and decay at rest. Sec.~\ref{sec:backgrounds} will give estimates for these backgrounds and we comment on their reducibility.

\subsubsection{Inelastic DM Transitions}
Models with non-minimal dark scalar (or fermion) spectra offer a unique signature to be exploited at an electron beam-dump experiment.
If excited states $\phi$ or $\psi$ decay promptly on scales $\lsim 10$ cm, then 
a unique handle on DM comes from the ground-states upscattering via the off-diagonal
gauge interactions in Sec.~\ref{sec:darkside} and transitioning into the short-lived excited states
\be
\chi  T \to \psi T  \to (\psi \to \chi~e^+ e^-) T
\ee
for fermions $\chi,\psi$. Similarly for scalars we have
\be
 \varphi  T \to \phi T  \to ( \phi \to \varphi~e^+ e^-) T,
\ee
where $T$ can be a target nucleus, nucleon, or electron. The detector signature of this process is a target recoil accompanied by an energetic $e^+$ $e^-$ pair.  This final state is difficult to mimic by a beam-originated or cosmic-originated background event.

%%%%%%%%%%%%%%%%%%%%%%%%%%%%%%%%%%%%%%%%%%%%%%%%
%%%%%%%%%%%%%%%%%%%%%%%%%%%%%%%%%%%%%%%%%%%%%%%%
%
% 					                SEC: Test Run Setup
%
%%%%%%%%%%%%%%%%%%%%%%%%%%%%%%%%%%%%%%%%%%%%%%%%
%%%%%%%%%%%%%%%%%%%%%%%%%%%%%%%%%%%%%%%%%%%%%%%%

\section{Test Run Setup}
\label{sec:testrun}

\begin{figure}[t!]
\includegraphics[width=8.6cm]{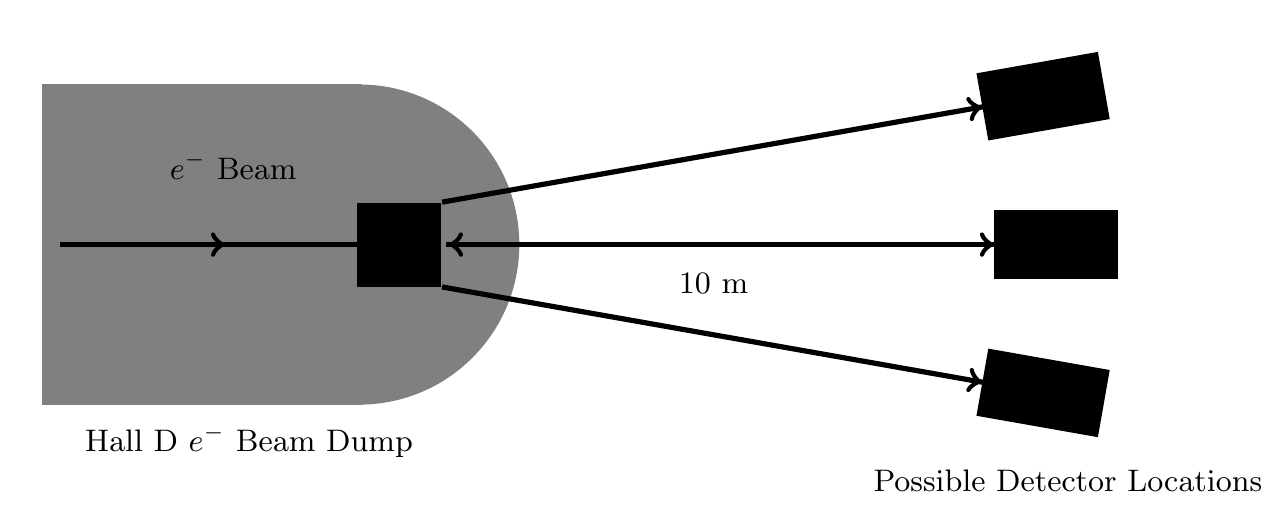}
\caption{  
Overhead view of the proposed experimental configuration behind Hall D. 
The detector can be placed either on-axis or displaced off-axis  to take
advantage of better acceptances for certain classes of DM particles. 
 }\label{fig:overhead}  %yes, ref in text
\end{figure}

In the test-run set up discussed in this paper, we assume placement of a small detector above ground roughly 10~m behind the electron beam-dump at JLab Hall D.  Fig.~\ref{fig:overhead} shows a schematic of possible test-run setups.  In a year of normal operations, Hall D will receive currents $\sim 200$ nA from CEBAF for a few months, which this experiment can use parasitically.  We therefore consider a benchmark of $10^{19}$ electrons on target (EOT) over a beam-on live time of 90 days.  The possibility of an off-axis detector is considered because the beamline into the dump is slightly below ground level.  An above-ground experiment would therefore be slightly misaligned with the beam axis.

%%%%%%%%%%%%%%%%%%%%%%%%%%%%%%%%%%%%%%%%%%%%%%%%
% 					           FIG:  Experimental Setup Figure 
%%%%%%%%%%%%%%%%%%%%%%%%%%%%%%%%%%%%%%%%%%%%%%%%

%%%%%%%%%%%%%%%%%%%%
%		 	Detector
%%%%%%%%%%%%%%%%%%%%

\subsection{Detector}
\label{sec:detector}
\begin{figure}[t!]
\vspace{0.4cm}
\includegraphics[width=8.8cm]{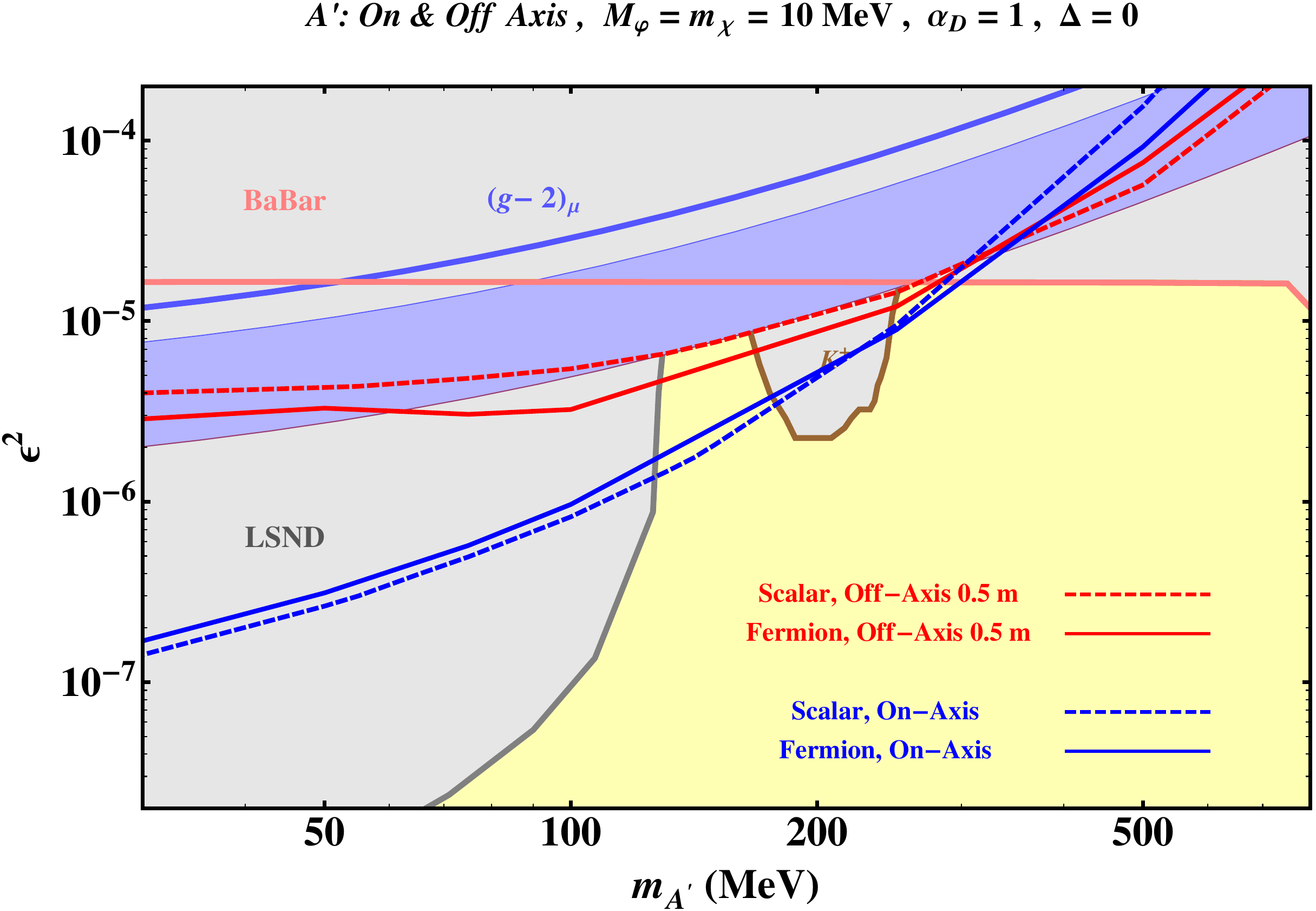} 
\includegraphics[width=8.8cm]{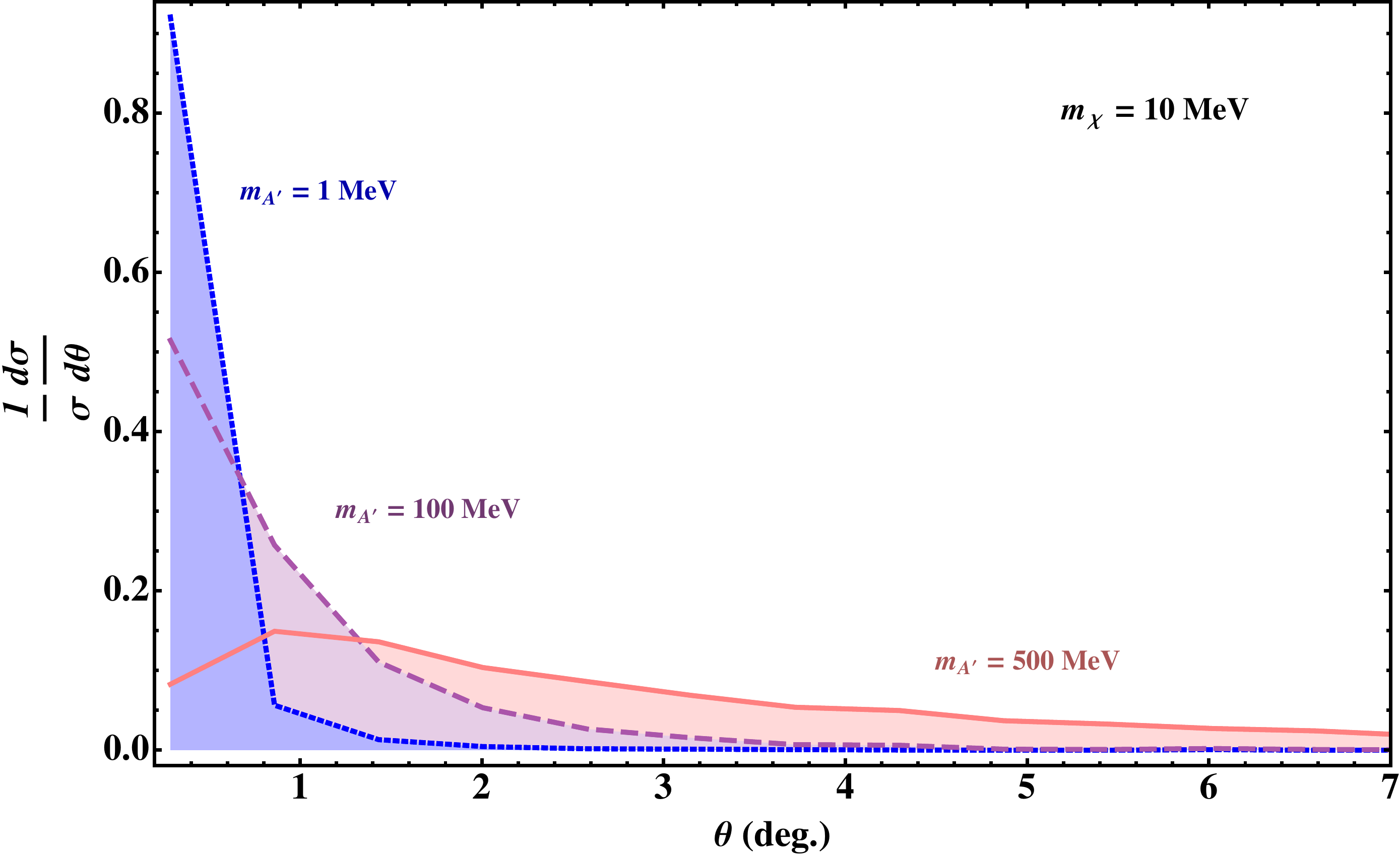} 
\includegraphics[width=8.8cm]{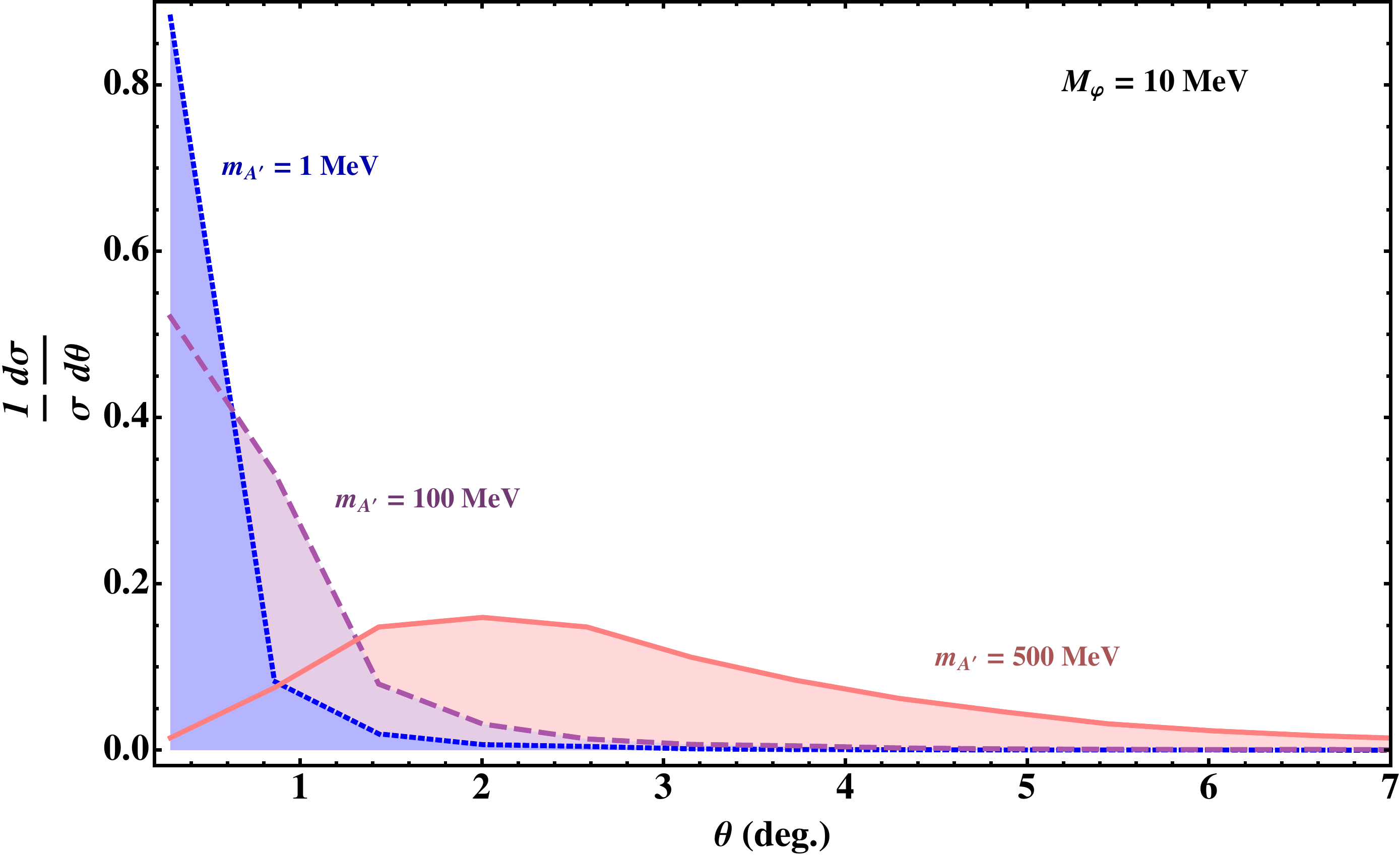}
\caption{
Top: yield comparison for scalar and fermion DM using both on and off-axis  detector positions (see Fig.~\ref{fig:overhead}).
Middle and bottom: angular distributions for 10 MeV fermion and scalar production respectively.
 Note the off-axis peak near $\theta \sim 2^o$, in the scalar distribution for $m_\apr = 500$ MeV. 
}\label{fig:kinematic}
\end{figure}

Inspired by the existing CORMORINO prototype \cite{MarcoSlides},  we
simulate DM-SM scattering in a 40 cm $\times$ 30 cm $\times$ 30 cm
detector of NE110 polyvinyltoluene ($C_{27} H_{30}$) plastic-scintillator. 
%This detector is small and mobile, so we consider both on-axis and off-axis 
%locations 10m  downstream of the beam dump. 
Fig.~\ref{fig:kinematic} shows the angular distribution 
with respect to the beam-line for various mediator masses for both fermion and scalar DM.

%%%%%%%%%%%%%%%%%%%%
%	     FIG Angular Distributions 
%%%%%%%%%%%%%%%%%%%%

%%%%%%%%%%%%%%%%%%%%
%		 Backgrounds
%%%%%%%%%%%%%%%%%%%%

\subsection{Backgrounds}
\label{sec:backgrounds}

The 12 GeV CEBAF at Jefferson Laboratory delivers electrons to experimental Halls A, B, C, and D. The proposed test-run in this article assumes that such an experiment would take place downstream of Hall D and follow the layout in Fig. \ref{fig:overhead}. Given the geography surrounding Hall D, a detector placed 10 meters behind the beam dump would be near or at ground level (the latter if run in off-axis mode). The various backgrounds associated with the test-run can be divided into two kinds: those originating from the beam, and those unrelated to the beam (cosmic-originated events). Beam-related backgrounds were estimated to be negligible even with $10^{22}$ EOT \cite{Izaguirre:2013uxa}, so we ignore these for the remainder of this section. In what follows we estimate the beam-unrelated backgrounds for leptophilic (electron channel) and inelastic models.  

Models where up-scattering to an excited state is followed by a prompt decay of the excited state leave a unique signature. The signal consists of an $e^+e^-$ pair, collectively depositing $\sim$ GeV energy, and a hard recoil from either an electron, nucleon, or nucleus.  If each of these particles could be separately resolved, then this signal would be easily separated from cosmogenic backgrounds.  For example, if the excited state lifetime is cm-scale, then the recoil and $e^+e^-$ pair would frequently appear in different cells of the detector.  Even for prompt decays and a simple plastic scintillator detector --- where the total energy deposited is probably the only observable signal --- this energy may be sufficient to stand out over backgrounds.  The same is true of the electron-scattering signal.  

The most important background process comes from cosmic muons which then decay to an electron. There are two possibilities to consider: stopped and decay-in-flight muons. The former can be removed entirely by vetoing on muon hits in a window as large as 100 $\mu$s and by cutting on $E_R>m_\mu\approx 52.5$ MeV. The timing window can be applied while still having little effect ($\sim 1\%$) on the detector livetime.

\begin{figure}[t!]
\includegraphics[width=8.6cm]{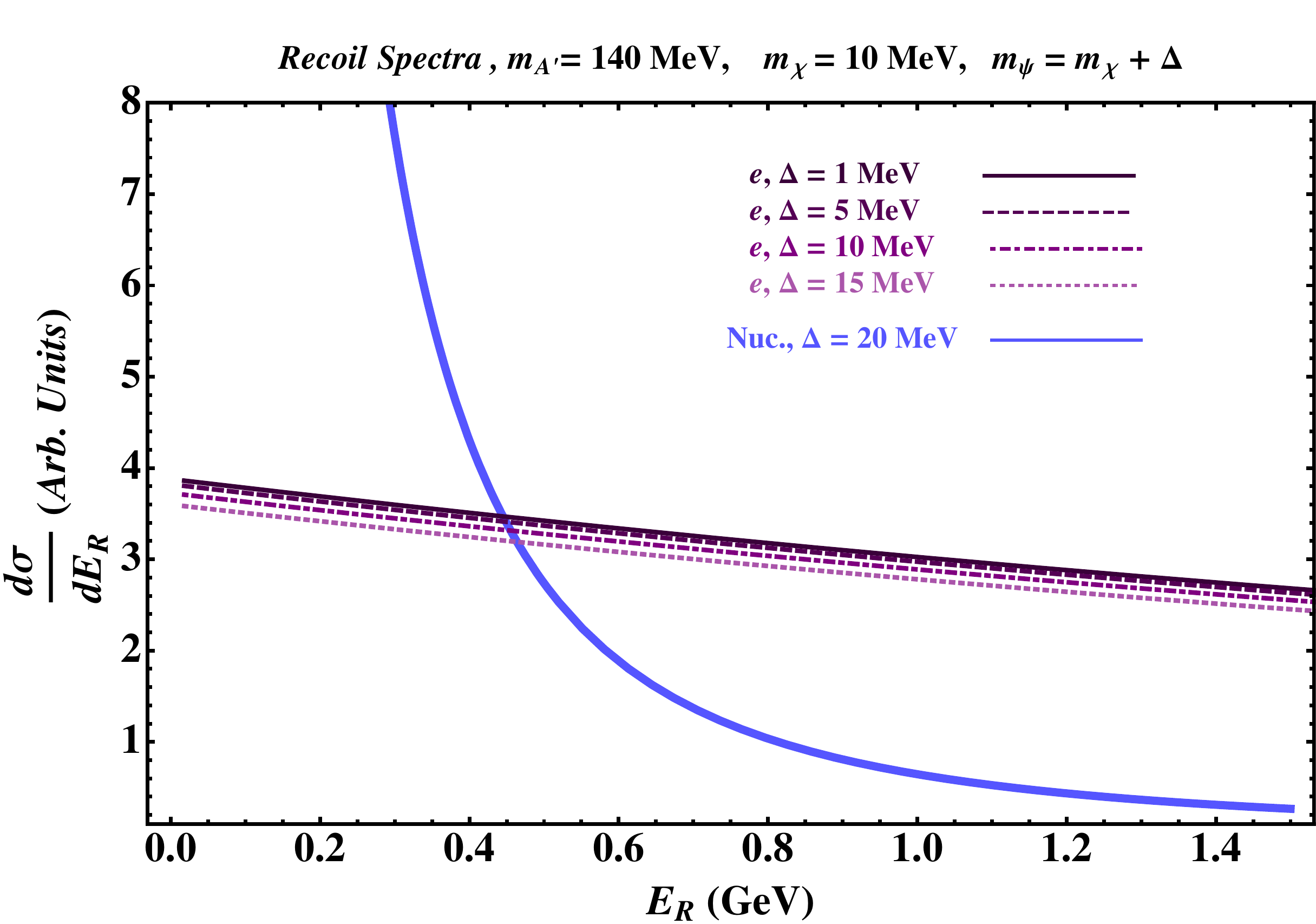}
\hspace{1.5cm}\includegraphics[width=8.5cm]{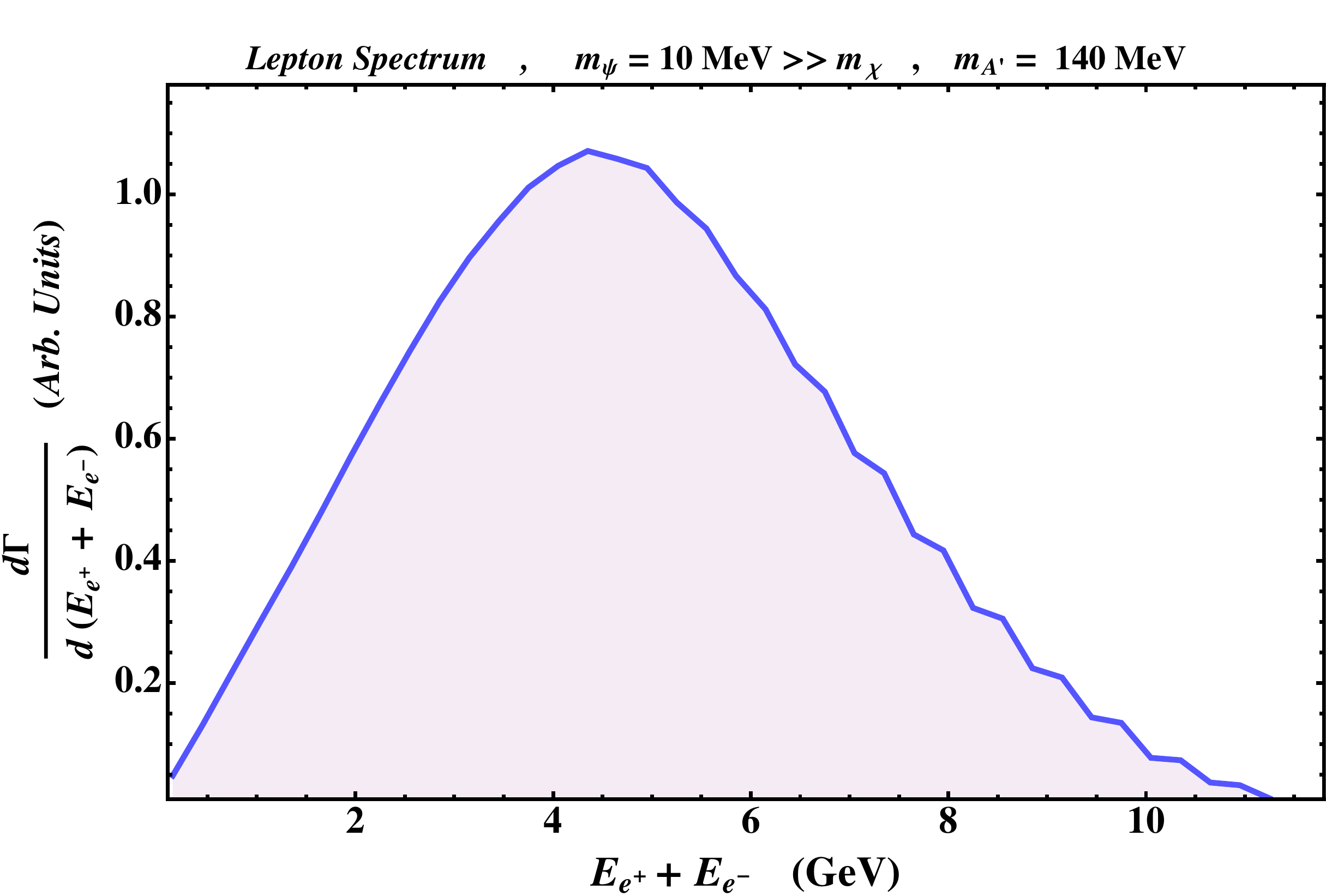}
\caption{Top: Differential recoil spectra in arbitrary units for fermion DM scattering inelastically off electrons  with different mass splittings $\Delta$ and a thick blue curve (color online) denoting 
the nucleon recoil distribution, whose shape does not change visually for the parameters we consider in this paper. Each
differential cross section is convolved with a monte carlo distribution of incoming $\chi$ energies that pass through the detector.
Bottom: Lab frame distribution of the combined electron and positron energies after $\psi \to \chi e^+ e^-$ de-excitation for various $\psi$ energies in the $m_{\psi} \gg m_{\chi}$ limit. Note that beam degradation (not simulated) would broaden the distribution and pull it towards lower energies.  Moreover, for $\Delta \lesssim m_\chi$ the peak energy scales as $\frac{E_{\text{beam}}}{2}\frac{\Delta}{m_\psi}$.
}\label{fig:recoildist}
\end{figure}

The rate of muon decays in flight within the detector can be inferred from measurements of the muon flux at sea-level  \cite{Kremer:1999sg}. For a CORMORINO-sized detector, we estimate a total rate of $\approx 10^{-2}$ Hz.  In 90 days of beam-on live time, this gives approximately $10^5$ decay-in-flight muons. While this background component is quite sizeable, it is also reducible with a high efficiency by vetoing events with electronic activity coincident with an incoming charged particle. 
Furthermore, most of the decaying muons are significantly less energetic than the the multi-GeV signals from electron recoils or de-excitation.  For example, requiring $E_\mu>2$ GeV reduces the decay-in-flight rate to $\approx 6\times10^{-4}$ Hz. Assuming a $10^3$ background-rejection efficiency yields $\mathcal{O}(10)$ decay-in-flight muon events in 90 days. In contrast, demanding requiring $E_R>2$ GeV has a weak (negligible) effect on the electron-recoil (inelastic de-excitation) signal efficiencies (See Fig.~\ref{fig:recoildist}). 
 
In Sec~\ref{sec:simulation} we discuss the details of the signal simulation. In Sec.~\ref{sec:Results} we give sensitivity estimates for the two classes of signals studied so far. These assume sensitivity at the 10-event level based on this estimates given above.  Though we do not explicitly model energy thresholds or beam degradation, these are expected to be at most $O(1)$ corrections to the signal yield. 

%%%%%%%%%%%%%%%%%%%%
%		 Simulation 
%%%%%%%%%%%%%%%%%%%%

\subsection{Simulation}
\label{sec:simulation}

The calculation of the signal yield is factorized into two reactions which are analogous to QED processes: production and re-scattering. On the production side, we use a modified version of 
\texttt{Madgraph 4} to simulate the process depicted in the top panel of Fig.~\ref{fig:scatfig} 
\be
e Z \rightarrow (A'^{(*)}\rightarrow\chi\bar\chi) e Z, 
\ee
where $Z$ stands for an individual nucleus in the beam dump target, made mostly of aluminum. A nuclear form factor from \cite{Kim:1973he} was used in the modified \texttt{Madgraph} version. The 
production simulation is used to extract the $dN/dE$, the energy profile for DM particles that pass
through the detector.   We do not model the effects of beam degeneration as it passes through the dump on $dN/dE$, but instead model only the production in the first radiation length of the dump.  The resulting signal yield is given by 
\be \label{eq:yield}
Y =   n_T \ell_D  \int_{ E_{c} }^\infty  \!\! dE_{R} \int_{E_{m} (E_R)  }^\infty   \!\! dE  \,
\frac{dN }{ dE  }  \frac{d\sigma}{dE_R} ~~,
\ee
for each scattering channel.  Here $n_T$ is the target particle density, $\ell_D$ is the longitudinal detector length, 
$E_{c}$ is the experimental cut on target recoils, $E$ is the incoming
DM energy,   $E_{m}(E_R)$ is the minimum energy for an incident particle
to induce a target recoil energy $E_R$, and $d\sigma/dE_R$ is the differential cross section
for a given channel -- see Appendix A.  

The detector reactions considered in this article are
depicted schematically in Fig. \ref{fig:scatfig} (bottom) for fermionic DM; analogous processes apply in the scalar scenario. 
Following  the notation of sections \ref{sec:fermions} and \ref{sec:scalars}, 
for the fermionic and scalar DM scenarios, the signatures of interest are 
\be
\chi ~T     &\to& (\psi \to \chi~e^+e^-) ~T ~~~, \\ 
 \varphi ~T &\to& (\phi \to \varphi~e^+e^-) ~T ~~, 
\ee
 where $T$ is a target nucleus (coherent scattering), nucleon (quasi-elastic scattering), or electron. In Fig.~\ref{fig:recoildist} (top) we show the 
 electron and nucleon recoil distributions for different values of  $\Delta$ using a monte carlo distribution of incoming DM energies that pass through the detector and (bottom) the lab frame
 $e^+ e^-$ energy distribution for different energies of the excited state.
 
 Unless otherwise specified, the recoil energy thresholds used in the analysis are 100 MeV for incoherent and electron scattering, and 100 keV for coherent nuclear scattering. 
For nuclear coherent scattering, a lower threshold is used to enhance the signal and get the $Z^2$ enhancement. In addition to a neutral current coherent scatter, one or more of the electrons from the decay of the excited state are required to scatter in the detector. This signature - an electron signal and a coherent scatter - renders a search for these classes of signals background-free.

%%%%%%%%%%%%%%%%%%%%%%%%%%%%%%%%%%%%%
%% Leptophilic Fig %%%
%%%%%%%%%%%%%%%%%%%%%%%%%%%%%%%%%%%%%

\begin{figure}[t!]
\vspace{0.1cm}
\includegraphics[width=8.6cm]{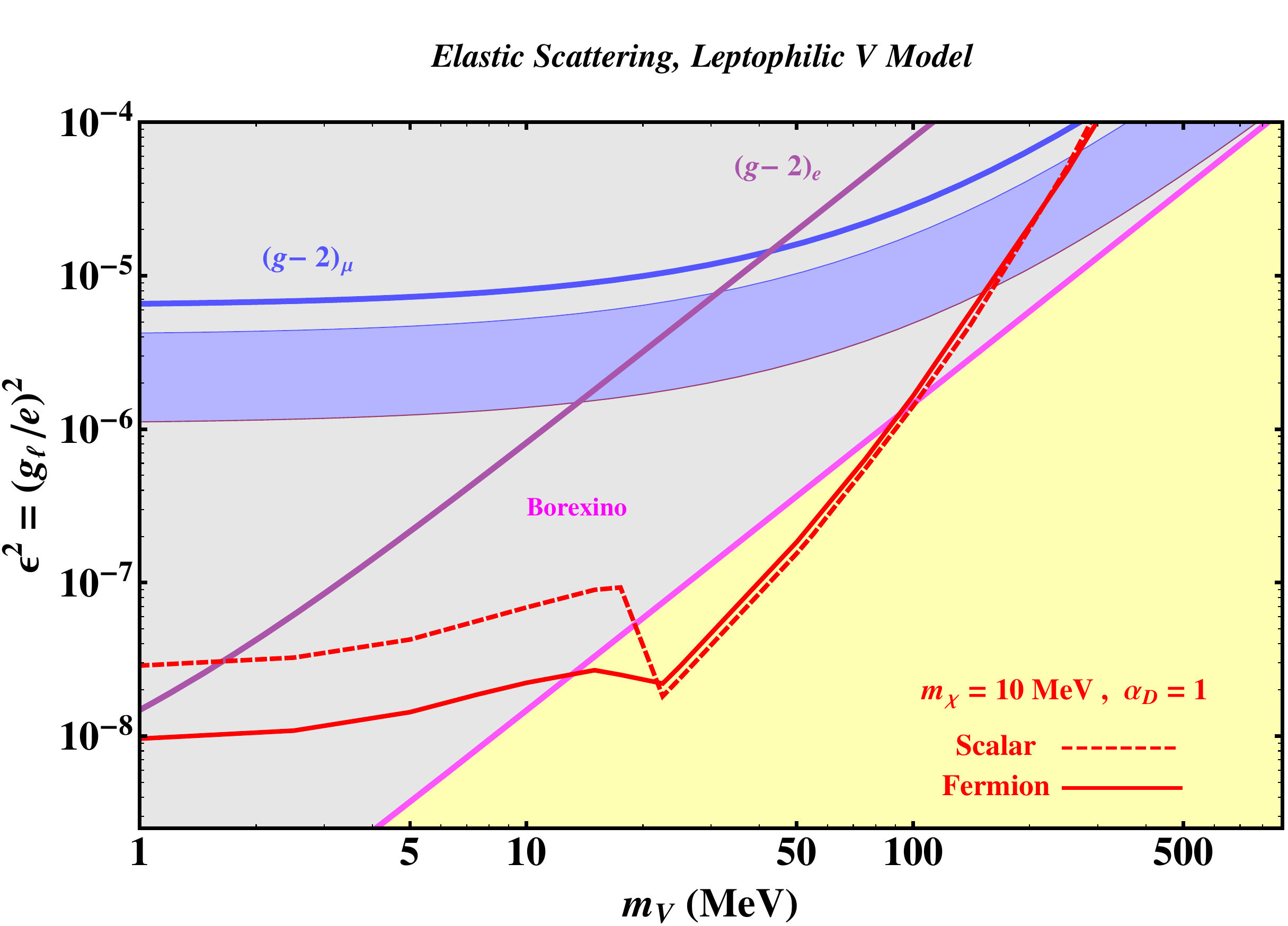}
\includegraphics[width=8.8cm]{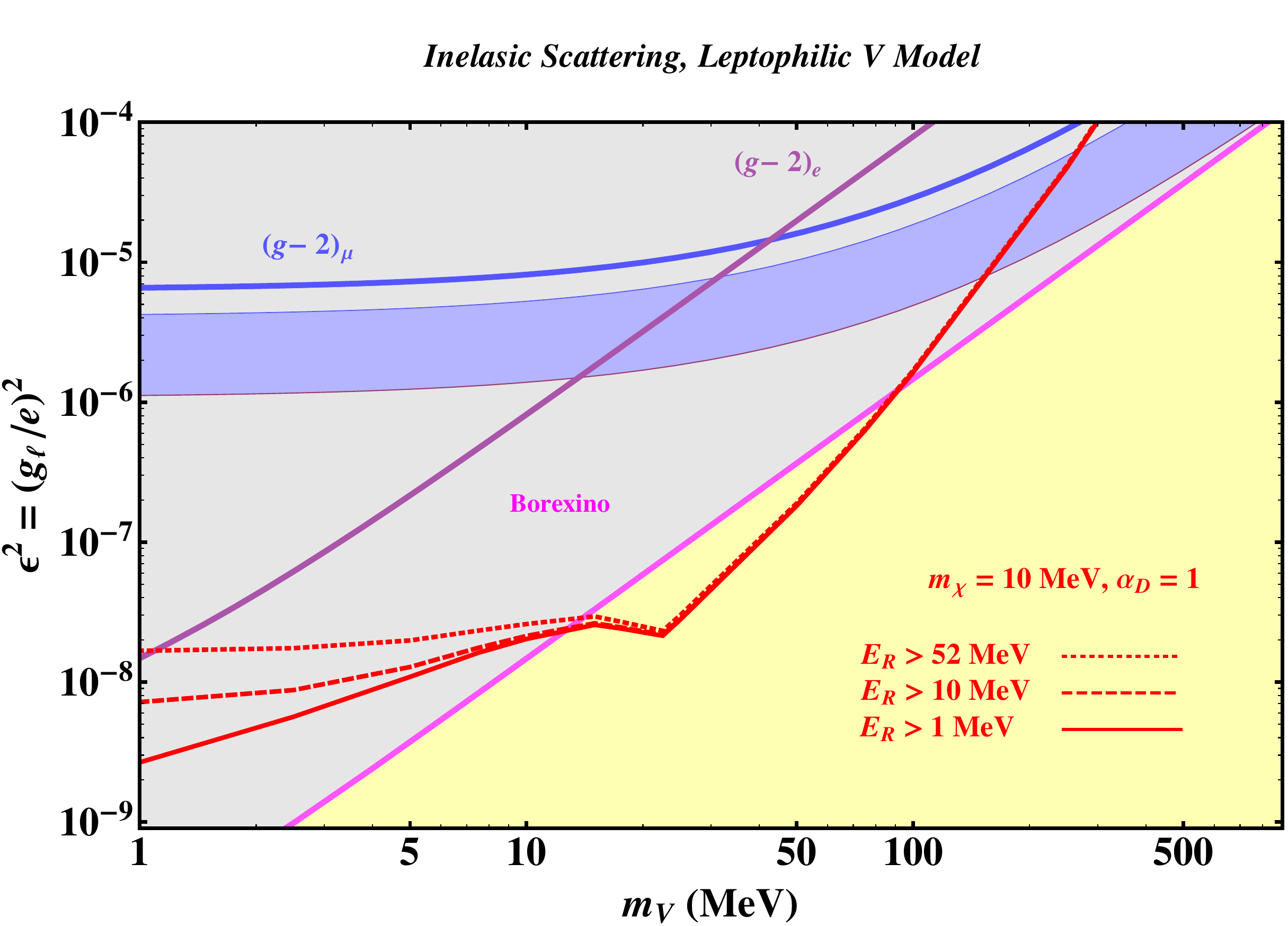}
 \caption{
  Top: solid and dashed red curves (color online) show 10 event contours for fermionic and scalar DM respectively
scattering off detector electrons  via leptophilic $V$ boson exchange (see Sec.~\ref{sec:leptophilic-mediator} and \ref{sec:darkside}). 
The Borexino constraint is extracted from \cite{Bellini:2011rx}.
Bottom: The 10-event sensitivity for different electron recoil energy thresholds.
 }
 \label{fig:leptophilic}
\end{figure}

\begin{figure}[htbp] \vspace{-0.1cm}
 \includegraphics[width=8.4cm]{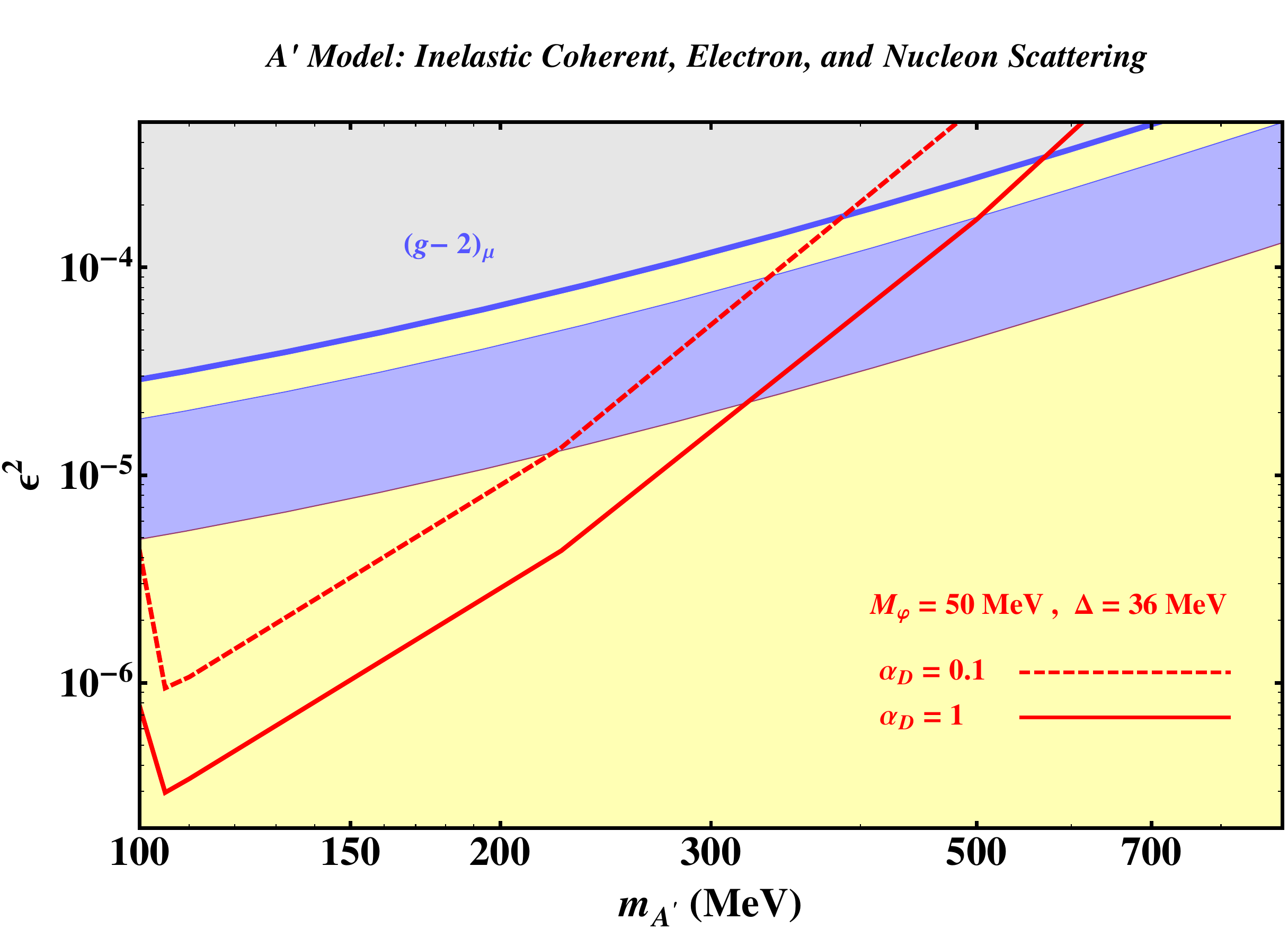} 
\includegraphics[width=8.4cm]{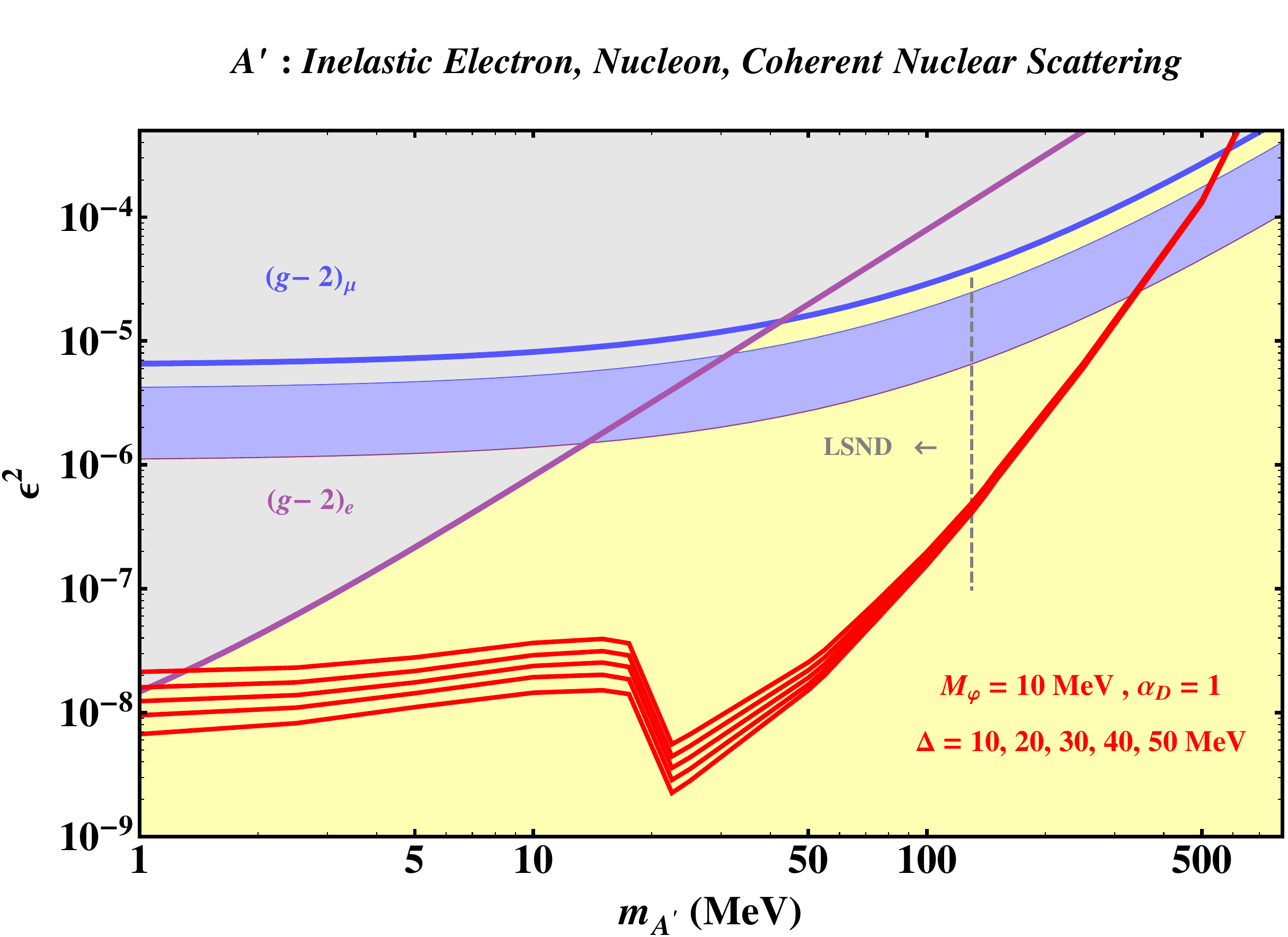}
\includegraphics[width=8.4cm]{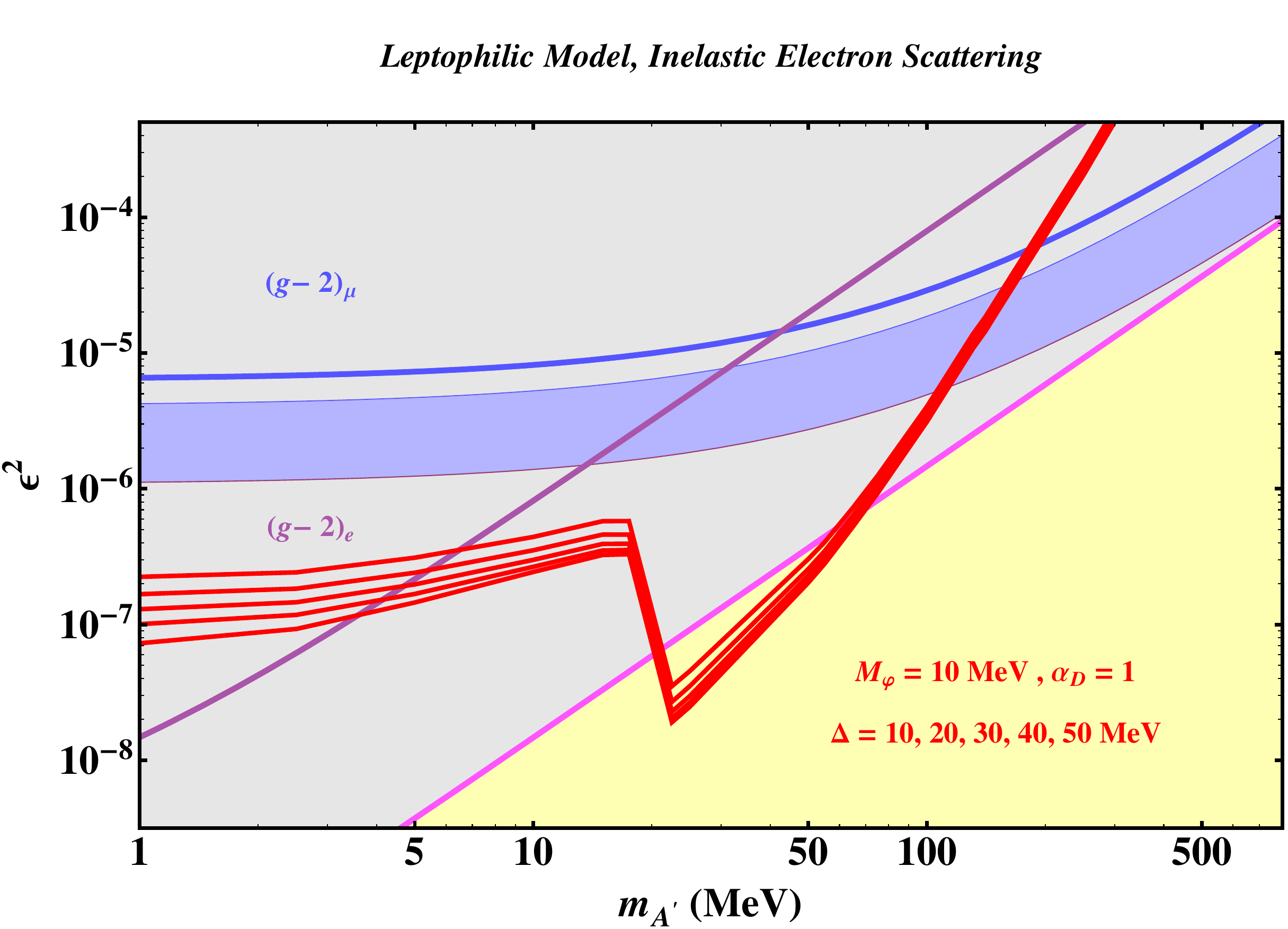}
  \caption{
  Scalar upscattering assuming  $\phi \to \varphi + e^+e^-$ deexcitation in the $\apr$ (top, middle) and leptophilic (bottom) scenarios.  
Top: $\alpha_D$ overlay for $M_\varphi  +  M_\phi = m_{\pi^0}$, which is inaccessible at LSND. 
Middle: $\apr$ mediated inelastic upscattering for various mass splittings. In this parameter space, there is a potential 
      constraint from LSND \cite{deNiverville:2011it} since electrons from the de-excitation can mimic 
    target-electron recoils inside the detector, but a full analysis is beyond the scope of this work.
The de-excitation signal is vetoed by the BaBar and rare $K^+$ decay searches shown in the middle plot of Fig. \ref{fig:depindep}.
         }
     \label{fig:inelastic2}
\end{figure}

 %
%In Fig.~\ref{fig:inelastic2} we show 10 event signal yields for scalar upscattering (top, middle) for the 
%combined electron, nucleon, and coherent-nuclear channels. Since a small-scale scintillator can only resolve bulk light-yield, 
%combining these channels is reasonable as information about the target recoil is difficult to separate from that of the $e^+e^-$ pair. 
% Fig.~\ref{fig:inelastic2} also shows an overlay of different $\Delta$ for the leptophilic scenario (bottom).
%For the details of our projections, see Appendix A.

%%%%%%%%%%
%Sec: Results
%%%%%%%%%

\section{Results}
\label{sec:Results}

The test-run set-up discussed in this paper can have discovery potential for new dark matter scenarios. In what follows, we discuss the sensitivity levels to the two smoking gun signals discussed in Sec.~\ref{sec:smoking}.

%%%%%%%%%%%%%%%%%%%%%%%%%%%%%%%%%%%%%%%%%%%%%%%%% 
%                           FIG   inelastic  (scalar) nucleon/coherent  (fermion) quasielastic
%%%%%%%%%%%%%%%%%%%%%%%%%%%%%%%%%%%%%%%%%%%%%%%%%

\subsection{Leptophilic scenario potential}
One kind of new physics that a test-run at JLab can be sensitive to is that of a leptophilic mediator between DM and the SM. 
Fig.~\ref{fig:leptophilic} shows the 10-event signal yields for electron scattering in the context of a leptophilic $A'$. The coupling between the DM and the $A'$ is given by $g_D$ and is assumed to be 1 for this scenario. Existing constraints, particularly those coming from solar neutrinos experiments already set strong bounds on the parameter space of this scenario. However, a full-scale experiment as discussed in \cite{Izaguirre:2013uxa} can cover significant new ground.

 %%%%%%%%%%%%%%%%%%%%%%%%%%%%%%%%%%%%%%%%%%%%%%%%% 
%                           FIG   inelastic  (scalar) nucleon/coherent  (fermion) quasielastic
%%%%%%%%%%%%%%%%%%%%%%%%%%%%%%%%%%%%%%%%%%%%%%%%%

\subsection{Inelastic transitions potential}
A small test run has particularly dramatic sensitivity to non-minimal dark sectors, where a DM excited state can decay  in the detector, depositing over a GeV of energy.   Both the ground ($\chi$ or $\varphi$ for fermion and scalar DM, respectively) and the excited states ($\psi$ and $\phi$) are produced in the beam-dump through an $A'$ radiated by an electron. For prompt excited states de-excitations, only the ground state makes it to the detector downstream of the beam dump, where it can up-scatter to the excited state. The latter then de-excites within the detector for certain regions of the parameter space. The top and middle plots in Fig.~\ref{fig:inelastic2}  show the 10-event level sensitivity at a test-run for the scalar DM scenario, for fixed choices of $\Delta = M_\phi - M_\varphi$. The $\Delta$ is chosen so as to have a prompt de-excitation within the detector. Thus, at least one of the $e^+$ $e^-$ pair is visible, regardless of whether the ground state $\varphi$ up-scatters off of a nucleus, nucleon, or electron in the detector, and regardless of the recoil energy. Note that B-factory and rare Kaon decay searches are insensitive to this scenario, because these analyses veto on extra event activity. Constraints from LSND for $m_A' < m_{\pi^0}$ do apply to this scenario, but are difficult to model; we simply indicate the kinematic limit to LSND sensitivity in Fig.~\ref{fig:inelastic2}(middle).  Fig.~\ref{fig:inelastic2} (bottom) shows similar projections varying $\Delta$ in the leptophilic scenario.

%%%%%%%%%%%%%%%%%%%%%%%%%%%
%%%%%%%%%%%%%%%%%%%%%%%%%%%
%
%                    		SEC: Conclusion
%
%%%%%%%%%%%%%%%%%%%%%%%%%%%
%%%%%%%%%%%%%%%%%%%%%%%%%%%

\section{Conclusion}
\label{sec:conclusion}
In this paper we have shown that a test run for a parasitic fixed-target experiment to search for DM 
at Jefferson Laboratory could have sensitivity to several 
 well motivated scenarios in only a few months of livetime. 
Motivated by efforts to launch such a test run experiment \cite{testProposal}, we  
considered signal yields for a small (sub meter-scale) plastic scintillator detector positioned above ground 10 meters downstream 
of a fixed target -- a geometry similar to that at the existing Hall D beam dump. 
With $10^{19}$ electrons on target, signal yields are sufficiently high to give a test run experiment 
unprecedented sensitivity to DM that couples to the visible sector through
 leptophilic mediators. The same experiment can also probe scenarios where the DM upscatters into an excited state.  In this case, the excited state's decay into $e^+e^-$ deposits GeV-scale energy in the detector, irrespective of the target electron, nucleus, or nucleon's recoil energy.  These signals can deposit considerably higher energies than the dominant cosmogenic backgrounds.  
 %\scratch{(we haven't fully analyzed the high energy cosmogenics)}
 These findings suggest that a small test-run demonstrating the viability of electron beam dump searches for light dark matter
 will provide new sensitivity to unexplored dark matter scenarios. 

%%%%%%%%%%%%%%%%%%%%%%%%%%%
%                     	 Acknowledgments 
%%%%%%%%%%%%%%%%%%%%%%%%%%%

\section*{Acknowledgments}
\medskip
We thank Brian Batell, Rouven Essig, John Jaros, Maxim Pospelov, Brian Shuve, and especially Haipeng An, Marco Battaglieri, Elton Smith, Stepan Stepanyan, 
and Raffaella De Vita for helpful conversations. The Perimeter Institute for Theoretical Physics is 
 is supported by the Government of Canada through Industry Canada
and by the Province of Ontario.
 
%%%%%%%%%%%%%%%%%%%%%%%%%%%
%%%%%%%%%%%%%%%%%%%%%%%%%%%
%
%                     Detector Scattering Appendix
%
%%%%%%%%%%%%%%%%%%%%%%%%%%%
%%%%%%%%%%%%%%%%%%%%%%%%%%%

\section*{Appendix A: Detector Scattering}
\label{sec:amps}
\renewcommand{\theequation}{A.\arabic{equation}}
\setcounter{equation}{0}

%%%%%%%%%%%%%%%%%%%%%%%%%%%
%                      Scalar Scattering Amplitude
%%%%%%%%%%%%%%%%%%%%%%%%%%%

\subsection*{Scalar Amplitude}
The amplitude for scattering $\varphi_1(p_1) + T(p_2) \to \varphi_2(k_1) + T(k_2)$ through a kinetically mixed photon is
\be
{\cal A} = \frac{\epsilon e g_D }{(t - m^2_{A^\prime})} \bar u(k_2) (\displaystyle{\not}{p_1} + \displaystyle{\not}{k_1} ) u(p_2)
\ee
where $t \equiv (p_2-k_2)^2 $ is the usual Mandelstam variable and 
$\varphi_i$ carries mass $m_i$. 
Squaring and averaging (summing) initial (final) state spins 
\be
|\overline{ \cal A}|^2
         &=& 
         \frac{2 (\epsilon e g_D)^2 }{(t - m^2_{A^\prime})^2} \biggl\{  
         (k_2\cdot p_1)(p_2\cdot p_1) + (k_2\cdot p_1)(p_2\cdot p_1)   \nonumber \\ 
         && \hspace{-1cm}  -  (k_2\cdot p_2)(p_1\cdot p_1) +  
        	(k_2\cdot p_1)(p_2\cdot k_1) + (k_2\cdot k_1)(p_2\cdot p_1)  \nonumber \\ 
	&&\hspace{-1cm}-  (k_2\cdot p_2)(p_1\cdot k_1)  +
         	(k_2\cdot k_1)(p_2\cdot k_1) + (k_2\cdot k_1)(p_2\cdot k_1)   \nonumber \\ 
	&&\hspace{-1cm}-  (k_2\cdot p_2)(k_1\cdot k_1) +         
         	(k_2\cdot k_1)(p_2\cdot p_1) + (k_2\cdot p_1)(p_2\cdot k_1)    \nonumber \\  &&\hspace{-1.cm} - (k_2\cdot p_2)(k_1\cdot p_1)   +          
    m_T^2 \left[     m_1^2  + m_2^2 +  2 (p_1\cdot k_1)  \right]  \biggr\}.~~~~~~~~
         \ee 

%%%%%%%%%%%%%%%%%%%%%%%%%%%
%                      Fermion Scattering Amplitude
%%%%%%%%%%%%%%%%%%%%%%%%%%%

\subsection*{Fermionic Amplitude}
The generic matrix element for fermion scattering $\chi_1(p_1) + T(p_2) 
\to \psi(k_1) + T(k_2)$ is 
\be
{\cal A} = \frac{\epsilon e g_D }{(t - m^2_{A^\prime})} [\bar u(k_2)\gamma^\mu u(p_2) ] [ \bar u(k_1)\gamma^\mu  u(p_1)]
\ee
where $\chi_i$ carries mass $m_i$. 
Squaring and averaging initial state spins 
\be
|\overline{ \cal A}|^2
 &=&
\frac{ 8(\epsilon e g_D)^2 }{(t - m^2_{A^\prime})^2}  \biggl[    
(k_1\cdot k_2)(p_1\cdot p_2) + (k_2\cdot p_1)(p_2\cdot k_1)  \nonumber \\ && \hspace{-0.5 cm}- 
 m_1 m_2 (k_2\cdot p_2) -  m_T^2 (p_1\cdot k_1) + 2 m_1 m_2 m_T^2  
\biggr] 
\ee

%%%%%%%%%%%%%%%%%%%%%%%%%%%
%                            Cross Section
%%%%%%%%%%%%%%%%%%%%%%%%%%%

\subsection*{Cross Section}
The differential cross section in the CM frame is 
\be \label{eq:xsecCM}
\frac{d\sigma}{d\Omega^*} &=&
\frac{  |\overline {\cal A}|^2}{  64 \pi^2 s  }    \frac{  |\vec k^*|  }{   \left|  \vec p^*  \right| } 
\ee
In terms of the lab frame recoil energy, the angular measure is $ d\cos \theta^* = (m_n/ |\vec p^*|  |\vec k^*| ) dE_R$,  
where the quantities  
\be
 |\vec k^*|^2 &=&  \frac{ (s-m_T^2-m_2^2)^2 - 4 m_T^2 m_2^2}{4s}   \\
 |\vec p^*|^2 &=&  \frac{ (s-m_T^2-m_1^2)^2 - 4 m_T^2 m_1^2}{4s} 
\ee
are the CM frame momenta for each particle in the initial and final state, respectively. 

If the target is a detector nucleus, there is additional form factor suppression,
so we modify the differential cross section with the replacement
\be
\frac{d\sigma}{dE_R} \longrightarrow  F(E_R)\frac{d\sigma}{dE_R}~~,
\ee
where, for momentum transfer  $q \equiv \sqrt{2 m_N E_R}$, the Helm form factor is \cite{Helm:1956zz, Duda:2006uk}
\be
F(E_R)  = \left(    \frac{3 j_1 (qr ) }{qr}     \right)^2    e^{-s^2 q^2 }  ~~, ~~   \\ 
r = \left( c^2 + \frac{7}{3} \pi^2 a^2  - 5 s^2 \right)^{1/2}~~, ~~ 
\ee
where $c = (1.23 A^{2/3} -0.6)$ fm,  $s = 0.9 ~{\rm fm}$ and $a = 0.52$ fm.

%%%%%%%%%%%%%%%%%%%%%%%%%%%
%                            Total Event Rate 
%%%%%%%%%%%%%%%%%%%%%%%%%%%

\subsection*{Total Event Rate}

For each target species $T$ in the detector (e.g. electrons, nucleons, or nuclei), the total event rate is formally
\be\label{eq:total-rate-appendix}
{\it Y} =   n_T \ell_D  \int_{ E_{R,c}} ^\infty  \!\! dE_{R} \int_{E_{m}(E_R)  }^\infty   \!\! dE \,
\frac{dN }{ dE  }  \frac{d\sigma}{dE_R}~~,
\ee
where $\ell_D$ is a characteristic detector length scale, $E_{R,c}$ is the experimental cut on recoil energies; inelastic kinematics require 
there to be a minimum recoil energy for a given splitting regardless of the cut, but
this is typically far below any feasible experimental cut. 
 The minimum incoming energy required for
 an incident particle of mass $m_1$ to scatter into a state of mass $m_2$
  for a fixed recoil energy $E_R$ 
\be
  E_{m} &=& 
\frac{     (E_{R}-m_T) \left[m_2^2  -m_1^2  + 2m_T (E_{R} - m_T  )   \right] + \sqrt{ \cal G}         }{ 4 m_T (E_{R} -m_T) } ~~, \nonumber \\ 
 && \hspace{-3cm}
\ee
where we define
\be
&&\!\!\!\!\!\!\!\!\!\!\!\!  {\cal G }  \equiv    
 (E_{R} -m_T )(E_{R} + m_T) \biggl\{  \bigl[            (m_1 -m_2)^2 + 2m_T(E_{R}  -m_T )    \bigr]   \nonumber \\  && \hspace{1.6cm} \times 
  \left[   (m_1 + m_2)^2 + 2m_T(E_{R}-m_T) \right]         \biggr\} ,  \!\!\!\!\!\!\!\!
 \ee
and the energy profile  $dN/dE$ is normalized to the number of DM particles passing through the detector. 

%%%%%%%%%%%%%%%%%%%%%%%%%%%
%%%%%%%%%%%%%%%%%%%%%%%%%%%
%
%                      3 Body Appendix 
%
%%%%%%%%%%%%%%%%%%%%%%%%%%%
%%%%%%%%%%%%%%%%%%%%%%%%%%%

\section*{Appendix B: Three Body  Decays}
\label{sec:amps2}
\renewcommand{\theequation}{B.\arabic{equation}}
\setcounter{equation}{0}

In this appendix we generalize the results from \cite{Dobrescu:2011px} and compute the de-excitation decays in the 
inelastic scenario where a DM particle up scatters into a heavier dark-sector state and
 promptly de-excites to a three-body final state inside the detector. 

%%%%%%%%%%%%%%%%%%%%%%%%%%%
%                            Fermion Decay
%%%%%%%%%%%%%%%%%%%%%%%%%%%
\subsection*{ Fermion Decay}
For fermions, the amplitude for de-excitation via $\psi(\ell_i) \to \chi(\ell_f)~e^+(p_+)  e^-(p_-)$ is 
\be
{\cal M }_\psi = \epsilon e g_D \frac{  [\bar u(p_2) \gamma^\mu u(p_1)][\bar u(p_+) \gamma_\mu v(p_-)]     }{2(p_+ \cdot p_-) - m_\apr^2} ~~, 
\ee
Squaring and summing spins, we have
\be
|{\cal \overline M}_\psi|^2 &=& \frac{16\, (\epsilon e g_D)^2 }{[m_\apr^2 - 2(p_+ \cdot p_-)  ]^2} \biggl[  (p_+\cdot p_2)(p_-\cdot \ell_i) \nonumber \\ && + 
 (p_+\cdot \ell_i)(p_-\cdot \ell_f) - m_2 m_1 (p_+ \cdot p_-)    \biggr]  ~~, ~
  \ee

%%%%%%%%%%%%%%%%%%%%%%%%%%%
%                           Scalar Decay
%%%%%%%%%%%%%%%%%%%%%%%%%%%
\subsection*{Scalar Decay} %factor of 2 relative to scatteringrate.pdf
For scalar decays, the three-body amplitude for $\phi(\ell_i) \to \varphi(\ell_f)~e^+(p_+)e^-(p_-)$ is 
\be
{\cal M }_\phi &=& \frac{ \epsilon e g_D  \, \bar u(p_+)( \displaystyle{\not}{\ell_i}+ \displaystyle{\not}{\ell_f})v(p_-)  }{2(p_+ \cdot p_-) - m_\apr^2} ~~,  \nonumber \\ 
&\simeq& \frac{ 2\epsilon e g_D  \, \bar u(p_+) \displaystyle{\not}{\ell_f} v(p_-)  }{2(p_+ \cdot p_-) - m_\apr^2} ~~. 
\ee
 Squaring and summing leptons spins yields
\be 
|{\cal \overline M}_\phi|^2 &=& \frac{16\,(\epsilon e g_D)^2   
 \bigl[       2  (p_+\cdot \ell_f)(p_-\cdot \ell_f) 
            -  m_1^2 (p_+ \cdot p_-)    \bigr] 
 }{(m_\apr^2 - 2p_+ \cdot p_-  )^2}      ~~,~~ \nonumber \\ \!\!\!\!\!\!\!\!\!
  \ee  

%%%%%%%%%%%%%%%%%%%%%%%%%%%
%                            Total Width
%%%%%%%%%%%%%%%%%%%%%%%%%%%
\subsection*{Total Width}
The width for both cases can be written
\be
\Gamma(\phi/\psi) = \frac{1}{(2\pi)^3 (8 m_{ \phi/\psi})} \int_0^{\,\Delta} \!\!\!     dE_+  
\int^{\,\varepsilon}_{E_+ -\varepsilon }  \!\!\!\!\! dE_-  |{\cal \overline A_{\phi/\psi}}|^2 ~~,~~~ 
\ee
where the parameter 
\be
\varepsilon \equiv \frac{\Delta - E_+}{ 1 - 2 E_+/m_{ \phi/\psi}}~~~,~~~
\ee
is the maximum energy of the final state $e^-$ for a fixed $E_+$. 
Using the kinematic identities in the limit $\Delta \ll m_{\psi, \phi}$
\be
p_+\cdot p_- &=& m_{ \phi/\psi} (E_+ + E_- - \Delta  ) \\
\ell_f \cdot p_\pm  &=& m_{ \phi/\psi} (\Delta - E_\mp) \\
\ell_i \cdot p_\pm &=& m_{ \phi/\psi} E_\pm
 \ee
we obtain 
\be
\Gamma(\phi \to \varphi ~ e^+  e^-) &=& \frac{4 \epsilon^2 \alpha \alpha_D \Delta^5}{15 \pi m_\apr^4}  + {\cal O}(\Delta^6)  ~,~~~~\\
\Gamma(\psi \to \chi ~e^+  e^-) &=& \frac{8 \epsilon^2 \alpha \alpha_D \Delta^5}{15 \pi m_\apr^4} 
 + {\cal O}(\Delta^6) ~,~~~~~~~
\ee
which confirm  Eqs. (\ref{eq:scalar-width}) and (\ref{eq:fermion-width}).
\vspace{0.5cm}
\subsection*{Decay-Signal Yield}
In addition to the target recoil yield, If the coupling to the $\apr$ is off-diagonal between different mass eigenstates, there
is a signal from the decay of the excited state inside the detector. Following the conventions in Eq.~(\ref{eq:total-rate-appendix}), 
the yield of de-excitation events is 
\be
{\it Y} =  n_T \ell_D  \int_{ E_{R,c}} ^\infty  \!\! dE_{R} \int_{E_{m}(E_R)  }^\infty   \!\! dE \,
{\cal \xi}(E, E_R) \frac{dN }{ dE  }  \frac{d\sigma}{dE_R}~,~~~~~~ \\ \nonumber
\ee
 where $ \xi \equiv \cal P F$ is an efficiency factor for which
\be 
{\cal P}(E, E_R) = 1 -  e^{- \ell_D/ \ell_{\phi,\psi}} ~~,~~
\ee
is the decay probability 
inside the detector, $ \ell_{\phi,\psi} \equiv c\gamma/ \Gamma_{\phi,\psi}$ is the decay length, and $\gamma$ is the decaying particle's boost factor in the lab frame. The  function 
\be
{\cal F}(E,E_R) \equiv \frac{1}{\,\,\,\Gamma_{\phi,\psi}} \int_{E^\pm_{\rm cut}}^{\gamma \Delta} dE_\pm \frac{d\Gamma_{\phi,\psi}}{dE_\pm}
\ee
ensures that only the visible fraction of decay byproduct is counted.
\bibliographystyle{apsrevM}
\bibliography{TestDump}
\end{document}